\documentclass[aps, prx,twocolumn,longbibliography,superscriptaddress,showpacs,amsmath,amssymb,floatfix]{revtex4}
\usepackage[latin9]{inputenc}
\usepackage{amssymb}
\usepackage{graphicx}
\usepackage{amsmath}
\usepackage{color}
\usepackage{mathrsfs}
\usepackage{float}
\usepackage{indentfirst}
\usepackage{mathrsfs}
\usepackage{float}
\usepackage{indentfirst}
\usepackage{textcomp}
\usepackage{comment}
\usepackage{mathtools}
\usepackage{natbib,hyperref}
\usepackage{soul}

\newcommand {\Y}{\textcolor {blue}}

\begin{document}

\title{Absence of spin liquid phase in the $J_1-J_2$ Heisenberg model on the square lattice}

\author{Xiangjian Qian}
\affiliation{Key Laboratory of Artificial Structures and Quantum Control (Ministry of Education),  School of Physics and Astronomy, Shanghai Jiao Tong University, Shanghai 200240, China}

\author{Mingpu Qin} \thanks{qinmingpu@sjtu.edu.cn}
\affiliation{Key Laboratory of Artificial Structures and Quantum Control (Ministry of Education),  School of Physics and Astronomy, Shanghai Jiao Tong University, Shanghai 200240, China}

\affiliation{Hefei National Laboratory, Hefei 230088, China}

\date{\today}

\begin{abstract}
We perform an in-depth investigation of the phase diagram of the $J_1-J_2$ Heisenberg model on the square lattice. We take advantage of Density Matrix Renormalization Group and Fully-Augmented Matrix Product States methods and reach unprecedented accuracy with large bond dimensions. We utilize excited-level crossing analysis to pinpoint the phase transition points. It was believed before that there exists a narrow spin liquid phase sandwiched by the N\'eel antiferromagnetic (AFM) and valence bond solid (VBS) phases. Through careful finite size scaling of the level crossing points, we find a direct phase transition between the N\'eel AFM and VBS phases at $J_2/J_1 = 0.535(3)$, suggesting the absence of an intermediate spin liquid phase. We also provide accurate results for ground state energies for a variety of sizes, from which we find that the transition between the N\'eel AFM and VBS phases is continuous. These results indicate the existence of a deconfined quantum critical point at $J_2/J_1 = 0.535(3)$ in the model. From the crossing of the first derivative of the energies with $J_2$ for different sizes, we also determine the precise location of the first order phase transition between the VBS and stripe AFM phases at $J_2/J_1=0.610(5)$. 
\end{abstract}

\maketitle

\begin{figure}[t]
	\includegraphics[width=70mm]{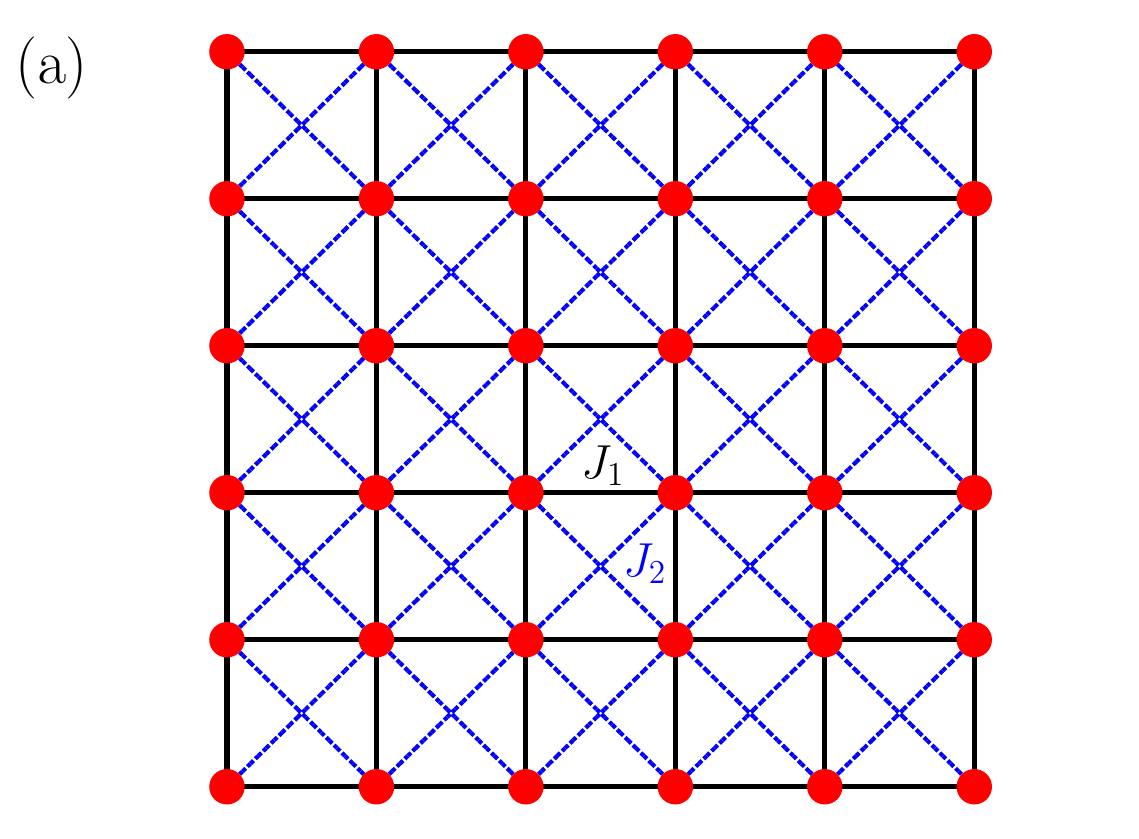}
	\includegraphics[width=80mm]{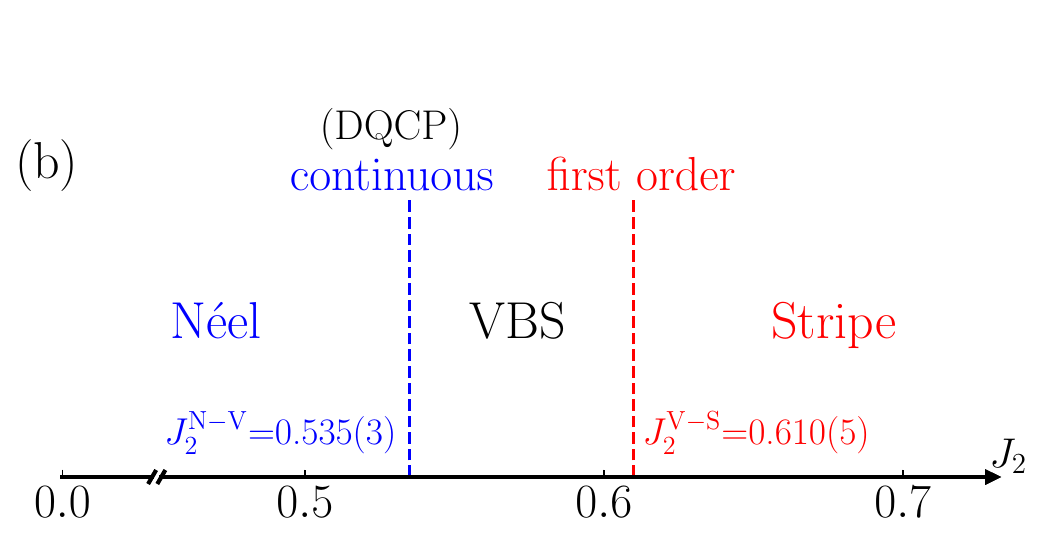}
   \caption{(a) The illustration of the $J_1-J_2$  Heisenberg model on the square lattice, with black and dashed blue lines indicating the
   nearest neighboring interaction $J_1$ and next-nearest neighboring interaction $J_2$, respectively. (b) The ground state phase diagram of the $J_1-J_2$ Heisenberg model on the square lattice with $J_2$ (we set $J_1=1$ as the energy unit). We find three different
   phases with the variation of $J_2$: N\'eel AFM, VBS, and stripe AFM. We find a direct continuous phase transition between the N\'eel AFM and VBS phases at $J_2^{\text{N-V}}=0.535(3)$, ruling out the existence of a quantum spin liquid phase between them, indicating the existence of a deconfined quantum critical point between the N\'eel AFM and VBS phases. At $J_2^{\text{V-S}}= 0.610(5)$, a first order transition occurs between the VBS and stripe AFM phases.}
   \label{phase}
\end{figure}

{\em \Y{Introduction --} }
The exploration of quantum phases and phase transitions in strongly correlated systems has long captivated the attention of physicists \cite{bookc,Xiao:803748,marino_2017}. Such investigations provide crucial insights not only into the fundamental behavior of matters but also into the exotic phenomena exhibited by these systems \cite{doi:10.1126/science.adc9487,Balents2010,Savary_2017,doi:10.1126/science.1091806,senthil2023deconfined,PhysRevLett.49.957,PhysRevLett.52.1583,PhysRevLett.53.722,PhysRevLett.128.146803,PhysRevLett.126.015301,Ludwig_2011,PhysRevLett.126.160604}. The $J_1-J_2$ Heisenberg model on the square lattice is a prominent example. The Hamiltonian of the model is
\begin{equation}
	H = J_1\sum_{\langle i,j \rangle}S_{i} \cdot S_{j}+J_2\sum_{\langle\langle i,j \rangle\rangle}S_{i} \cdot S_{j}
\end{equation}
where $S_i$ is the spin-1/2 operator on site $i$, and the summations are taken over nearest-neighbor ($\langle i,j \rangle$) and next-nearest-neighbor ($\langle \langle i,k \rangle \rangle$) pairs, as shown in Fig.~\ref{phase} (a). 
This model, which comprises nearest neighboring ($J_1$, the energy unit in this work) and next nearest neighboring ($J_2$) exchange interactions, exhibits a delicate interplay between competing magnetic interactions. This competition makes the model a well-known playground to search for exotic quantum states (quantum spin liquids (QSL), for example) other than on geometrically frustrated lattice as kagome lattice \cite{doi:10.1126/science.1201080,PhysRevLett.109.067201,PhysRevB.87.060405,PhysRevLett.118.137202,PhysRevX.7.031020}. 
The understanding of this kind of exotic states may serve as a critical piece in unraveling the enigmatic behavior of doped Mott materials and high-temperature superconductivity \cite{RevModPhys.78.17,PhysRevLett.127.097002,doi:10.1073/pnas.2109978118,PhysRevB.102.041106,xu2023coexistence,lu2023sign,qu2023dwave}.

Over the past three decades, enormous research efforts have been dedicated to the exploration of the phase diagram of the $J_1-J_2$ Heisenberg model on the square lattice. This model has become a focal point in the study of quantum magnetism. 
It has become evident that distinct magnetic orders emerge at different regimes of the parameter space.
In the zero $J_2$ limit, the ground state displaces N\'eel antiferromagnetic (AFM) order \cite{PhysRevB.56.11678} which stretches to a finite $J_2$ value. In the opposite limit where $J_2$ is infinitely large, the model decouples into two isolated Heisenberg models on the two sublattices of the original square lattice. A large but finite $J_2$ couples these two sublattices and the ground state is known to have the so-called stripe AFM order \cite{PhysRevLett.63.2148,QMC_Imada}.   
In the intermediate range, roughly encompassing $0.5 \lesssim J_2 \lesssim 0.6$, an intriguing nonmagnetic regime emerges. These insights have been gleaned through a multifaceted approach, including exact diagonalizations \cite{PhysRevLett.63.2148,PhysRevB.43.10970,refId0,PhysRevB.74.144422}, series expansions \cite{PhysRevB.54.3022,PhysRevB.60.7278,PhysRevB.73.184420}, density-matrix renormalization group (DMRG) \cite{PhysRevB.86.024424,PhysRevLett.113.027201,PhysRevLett.121.107202}, (infinite) projected entangled-pair state ((i)PEPS) \cite{PhysRevB.97.174408,10.21468/SciPostPhys.10.1.012,LIU20221034,PhysRevB.94.075143,PhysRevB.105.195140}, neural network \cite{PhysRevX.11.031034}, and Quantum Monte Carlo (QMC) \cite{PhysRevB.88.060402,QMC_Imada,PhysRevB.102.014417}.

However, despite the wealth of results and analyses, the nature of the nonmagnetic regime in the $0.5 \lesssim J_2 \lesssim 0.6$ range remains a subject of intense debate and active research. Within this intriguing region, several competing states have been found with a variety of methods. These states include columnar Valence Bond Solid (VBS) \cite{PhysRevLett.65.1072,PhysRevLett.66.1773,PhysRevB.60.7278}, plaquette VBS \cite{PhysRevLett.84.3173,PhysRevB.74.144422,PhysRevLett.113.027201,PhysRevB.85.094407}, as well as quantum spin liquids with \cite{PhysRevB.86.024424} or without \cite{PhysRevB.88.060402,PhysRevLett.121.107202,PhysRevX.11.031034,LIU20221034} spin gap.

Much of the controversy surrounding this topic can be attributed to the difficulty of simulating large systems with enough accuracy and the challenges posed by finite size scaling of the order parameters or gaps, which could introduce large uncertainties near the critical points \cite{K_Nomura_1995,PhysRevB.104.075142,PhysRevB.84.224435,OKAMOTO1992433,PhysRevLett.104.137204}.
To address these issues, we adopt the level-spectroscopy approach, in which the finite-size transition points are determined through the identification of excited-level crossings \cite{PhysRevLett.121.107202,K_Nomura_1995,OKAMOTO1992433,PhysRevB.54.R9612,PhysRevLett.104.137204,PhysRevLett.115.080601,10.1063/1.3518900,PhysRevB.94.144416}. This approach was widely adopted in very recent studies of the same model studied in this work and other models \cite{PhysRevLett.121.107202, PhysRevX.11.031034,PhysRevB.105.L060409,Wang_2022}. The smooth size dependence exhibited by these crossing points allows for more reliable extrapolations to thermodynamic limit, surpassing the limitations encountered in past studies relying solely on order parameters, thus allowing us to accurately determine the phase boundaries.

In this Letter, we perform an in-depth investigation of the $J_1-J_2$ Heisenberg model on the square lattice using state-of-the-art numerical techniques, including DMRG \cite{PhysRevLett.69.2863,PhysRevB.48.10345,RevModPhys.77.259} and the recently developed FAMPS methods \cite{Qian_2023}. We take advantage of the $SU(2)$ symmetry in the calculations and reach bond dimension (kept states) to as large as $15000$ SU(2)
multiplets, which contains about $60000$ U(1) states.  For FAMPS \cite{Qian_2023}, we preserve up to equivalently $22000$ U(1) states \cite{foot1}. We obtain accurate results on cylinders with width up to $14$ by careful extrapolations with truncation errors in both DMRG and FAMPS. By utilizing the level-spectroscopy combined with reliable finite size scaling, we find a direct phase transition between N\'eel AFM and VBS phases at $J_2^{\text{N-V}} = 0.535(3)$ -- in contrast to the previous predictions of a QSL phase between them. We observe no tendency of singularity in the first and second derivative of ground state energy with respect to $J_2$ at this transition point, indicating the transition between the N\'eel AFM and VBS phases is continuous. This evidence implies this transition is a deconfined quantum critical \cite{doi:10.1126/science.1091806,senthil2023deconfined} type. The characterization of this transition deserves further investigation. We also determine the precise location of the first order phase transition between the VBS and stripe phases at $J_2^{\text{V-S}} = 0.610(5)$ from the crossing of the first derivative of the energies with $J_2$ for different
sizes. An illustration of the phase diagram is shown in Fig.~\ref{phase} (b).

{\em \Y{Methods --} }   
DMRG is now arguably the workhorse for the accurate simulation of one-dimensional and quasi-one-dimensional quantum systems \cite{PhysRevLett.69.2863,PhysRevB.48.10345,RevModPhys.77.259}. As a variational method, the wave-function ansatz of DMRG is known as Matrix Product states (MPS) \cite{PhysRevLett.75.3537,PhysRevB.73.094423,doi:10.1080/14789940801912366}, which is defined as
\begin{equation}
 |\text{MPS} \rangle=\sum_{\{\sigma_i\}} \text{Tr}[A^{\sigma_1}A^{\sigma_2}A^{\sigma_3}\cdots A^{\sigma_n}]|\sigma_1 \sigma_2 \sigma_3\cdots \sigma_n\rangle
 \end{equation}
 where $A$ is a rank-3 tensor with one physical index $\sigma_i$ with dimension $d$ and two auxiliary indices with dimension $D$. 
 DMRG has also been widely used in the study of two-dimensional quantum systems with narrow cylinder geometries \cite{PhysRevLett.127.097002,doi:10.1073/pnas.2109978118,xu2023coexistence,lu2023sign,qu2023dwave}.

 In the pursuit of even higher accuracy and the alleviation of the entanglement limitation in the simulation of wider systems with DMRG, we resort to a recently developed method named FAMPS \cite{Qian_2023}. FAMPS is an extension of DMRG by adding an additional layer of tensors known as disentanglers \cite{foot3,PhysRevLett.101.110501} connecting to the physical indices of DMRG. It is defined as
 \begin{equation}
	|\text{FAMPS}\rangle=D(u)|\text{MPS}\rangle
\end{equation}
where $D(u)=\prod_{m} u_m$ denotes an additional disentangler layer. This extension empowers FAMPS with the extraordinary capability to produce more accurate results for wider quantum systems, while maintaining the computational efficiency (O($D^3$)) \cite{PhysRevLett.126.170603,Qian_2023,PhysRevB.105.205102} of DMRG with small overhead (O($d^4$)). 

Moreover, we have invested efforts into optimizing the code efficiency.
Through techniques such as parallelization 
and the exploitation of SU(2) symmetry \cite{I.P.McCulloch_2002,WEICHSELBAUM20122972}, we are able to push the kept states which determine the accuracy of the simulation in DMRG and FAMPS to unprecedented value, $60000$ ($22000$) U(1) states for DMRG (FAMPS), setting the new limit of the numerical simulation.

We employ these state-of-the-art numerical techniques on a diverse set of $L\times 2L$ cylinder systems, spanning a range of sizes from $L=6$ to $L=14$ and $J_2$ values from $J_2=0.45$ to $J_2=0.65$, and giving well-converged results with extrapolation with truncation errors in DMRG and FAMPS calculations. 

As mentioned earlier, we utilize the level-crossing of excited states with different quantum numbers to determine the phase transition points. For excitations in the $S = 0$ sector, we target multiple states \cite{PhysRevB.48.10345,li2023accurate,baker2023direct} in the $S = 0$ subspace since the ground state lies in the $S = 0$ sector. For excitations in $S = 1$ and $S = 2$ sectors, we obtain the energies by performing the calculation in the desired subspace to obtain the excited state.

\begin{figure}[t]
   \includegraphics[width=80mm]{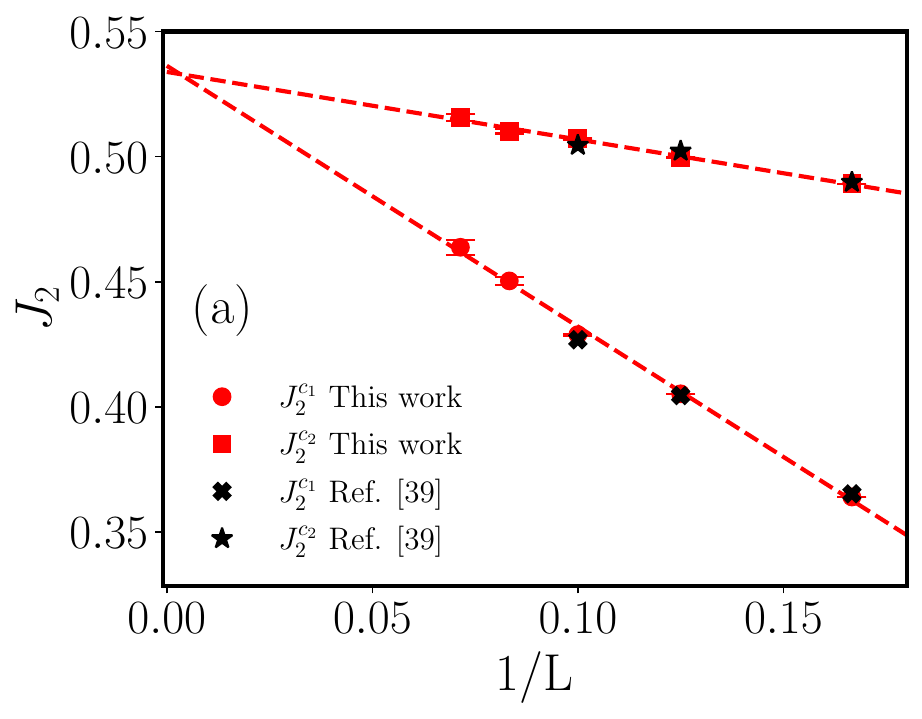}
   \includegraphics[width=38mm]{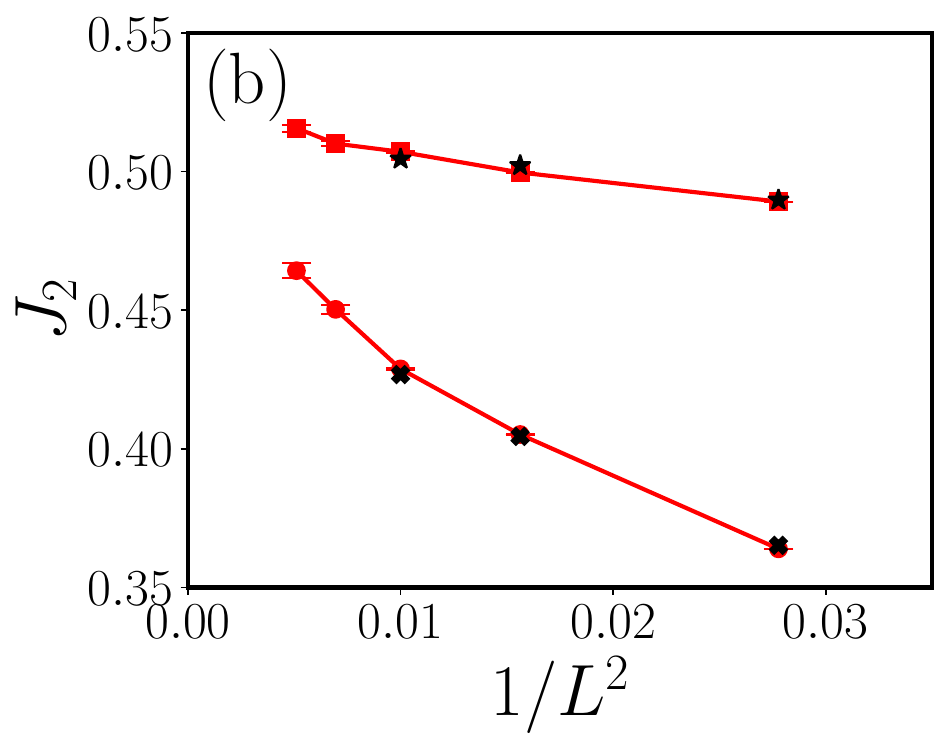}
   \includegraphics[width=38mm]{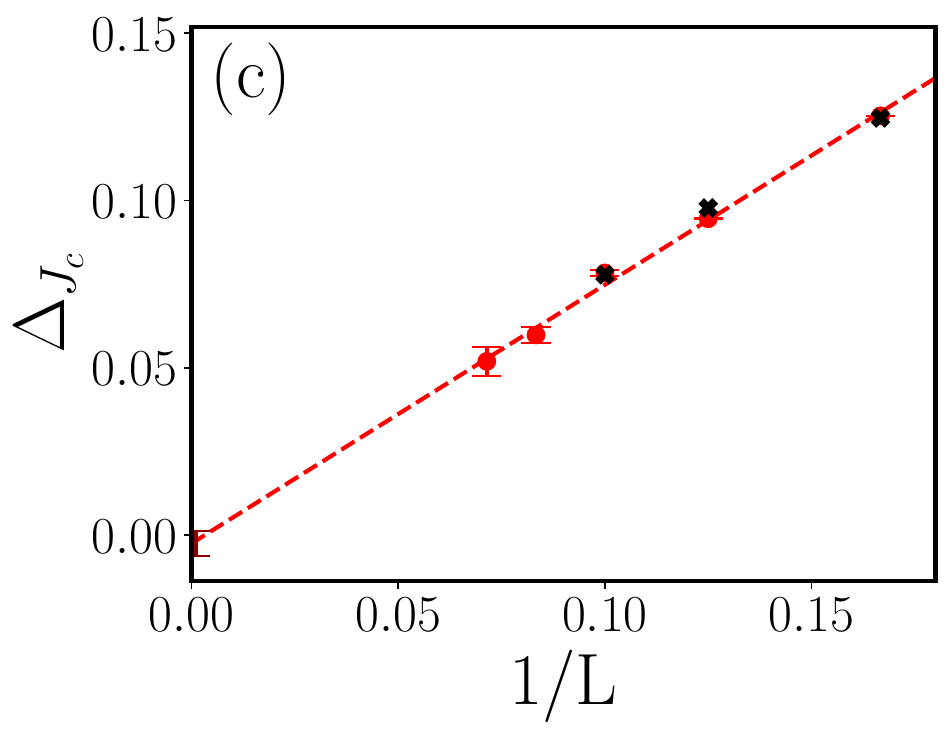}
  \caption{(a) The excited-level crossing points $J_2^{c_1}$ and $J_2^{c_2}$ as a function of $1/L$. The singlet-quintuplet crossing point $J_2^{c_1}$ is interpreted as N\'eel AFM phase boundary and singlet-triplet crossing point $J_2^{c_2}$ is identified as the onset of VBS phase \cite{PhysRevLett.121.107202}. We also include the results from Ref \cite{PhysRevLett.121.107202} which is consistent with our results. The extrapolated critical points by linear fits are $J_2^{c_1}=0.537(4)$, $J_2^{c_2}=0.533(1)$, providing strong evidence in favor of a direct transition between the N\'eel AFM and VBS states, ruling out the presence of an intermediate QSL phase. {(b) We plot the same
  data in (a) by changing x axis to $1/L^2$. We can clearly see the deviation from a straight line of the data. (c) The difference of the two crossing point $\Delta_{J_c}=J_2^{c_2}-J_2^{c_1}$ vs $1/L$. We can clearly see that $\Delta_{J_c}$ scales linearly with $1/L$ and vanishes in the thermodynamic limit.}
  }
  \label{crossing}
\end{figure}

\begin{figure}[t]
	\includegraphics[width=42mm]{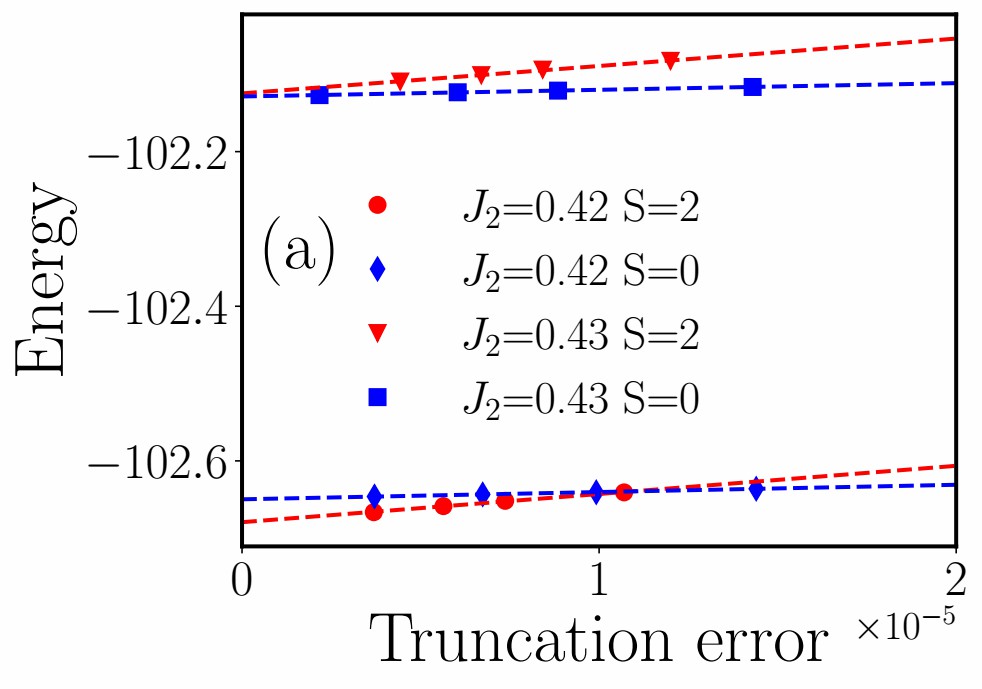}
	\includegraphics[width=42mm]{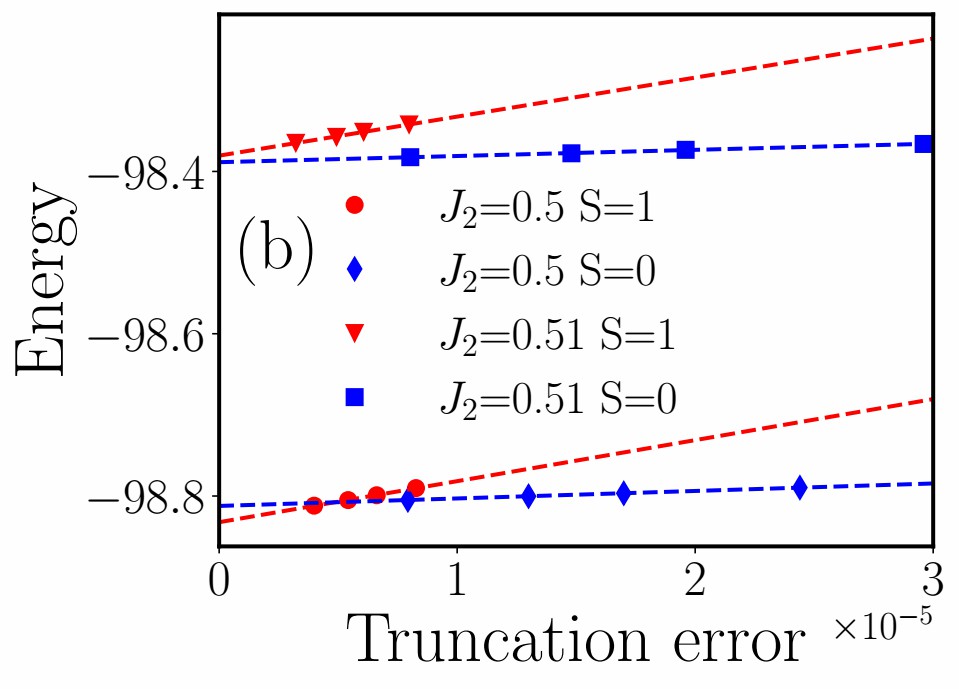}
	\includegraphics[width=42mm]{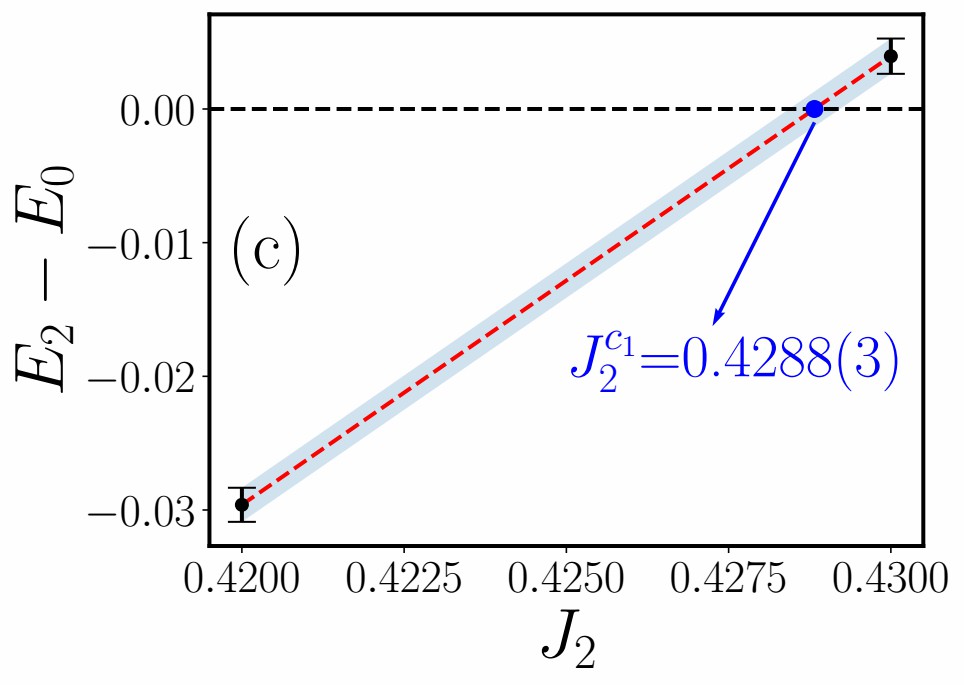}
	\includegraphics[width=42mm]{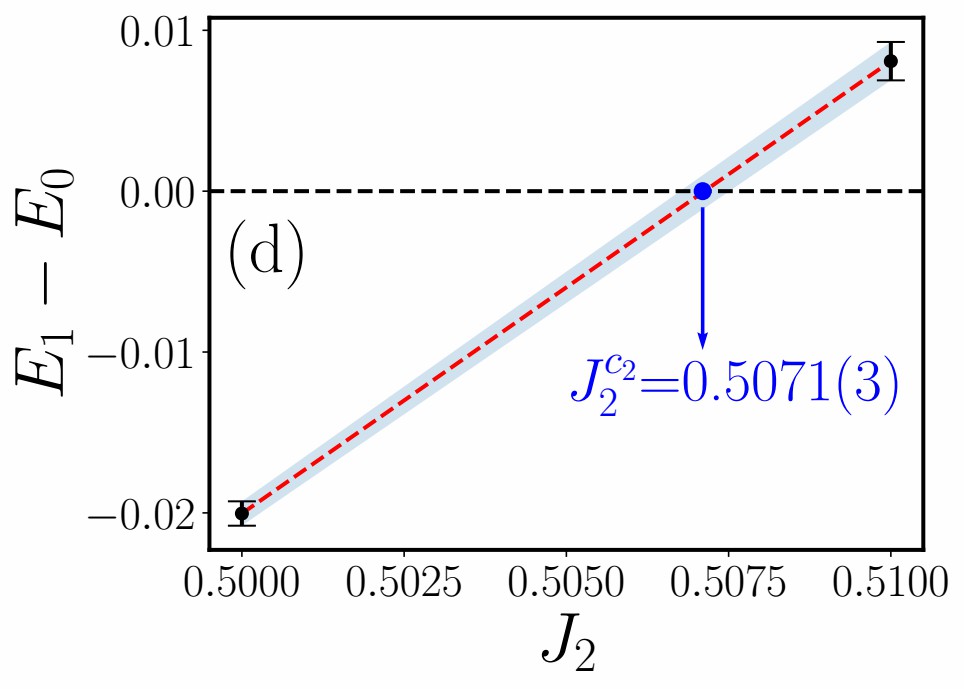}
	   \caption{Energies of the excitations in the $S=0, 1$, and $2$ sectors for a few $J_2$ values of $10\times 20$ cylinder system calculated by DMRG. In (a) and (b), DMRG energies are plotted as a function of truncation errors. We keep maximally equivalent $40000$ U(1) states in these DMRG calculations to ensure the convergence of the results. The excited state in the $S=0$ sector is calculated with the multi-target states DMRG algorithm \cite{PhysRevB.48.10345,li2023accurate,baker2023direct}. (c), (d) shows interpolation procedure to determine the excite-level crossing points for singlet-quintuplet and singlet-triplet excitations, respectively.}
	   \label{demo_crossing}
   \end{figure}

{\em \Y{Level-spectroscopy --} }
We begin our investigation by studying the region between the N\'eel AFM and the VBS phases. 
Historically, the regime encompassing $0.5 \lesssim J_2 \lesssim 0.6$ has been a focal point of considerable debate and intensive research. Numerous investigations, predominantly reliant on order parameters, have sought to elucidate the distinct phases characterizing this intricate region. Recent developments have resulted in a growing consensus that the emergence of the VBS state occurs within this region with onset $J_2$ varying from $0.52$ to $0.56$. \cite{PhysRevX.11.031034,LIU20221034,PhysRevX.12.031039,PhysRevLett.113.027201}. However, a contentious issue lingers concerning the potential possibility of a QSL phase at smaller values of $J_2$ (approximately around $J_2 \approx 0.5$) \cite{PhysRevB.97.174408,10.21468/SciPostPhys.10.1.012,PhysRevX.11.031034,PhysRevLett.113.027201,PhysRevB.94.075143}. This controversy primarily arises from the finite-size scaling of order parameters or gaps, which could introduce large uncertainties near the critical points, thus challenging the precise determination of critical points and phase boundaries within this regime \cite{OKAMOTO1992433}.

In Ref \cite{PhysRevLett.121.107202,K_Nomura_1995,OKAMOTO1992433,PhysRevB.54.R9612,PhysRevLett.104.137204,PhysRevLett.115.080601,10.1063/1.3518900,PhysRevB.94.144416}, the authors introduced a numerical level-spectroscopy method, wherein finite-size transition points are identified through excited energy crossings. This approach is rooted in the fundamental understanding that quantum phases are distinguished by their distinctive characteristics within excitation spectra. In finite-sized systems, low-lying excitations bear distinct quantum numbers corresponding to different phases, rendering them invaluable probes for the detection and characterization of phase transitions. This innovative approach is known to have a smooth finite-size scaling \cite{OKAMOTO1992433}, allowing us to accurately determine the phase boundary. This method was also adopted in recent studies of the same model studied in this work \cite{PhysRevLett.121.107202, PhysRevX.11.031034}. Specifically, the crossing point between the singlet and quintuplet excited states, denoted as $J_2^{c_1}$, is interpreted as the N\'eel AFM phase boundary, and the crossing point between the singlet and triplet excited states, marked as $J_2^{c_2}$, is identified as the onset of the VBS phase \cite{PhysRevLett.121.107202}. 

In Ref \cite{PhysRevLett.121.107202}, the authors determined the phase diagram by DMRG with the level-spectroscopy method. However, due to the limited system sizes calculated in Ref \cite{PhysRevLett.121.107202}, it is not easy to determine the actual finite size scaling behavior of the excited-level crossing points. 

Our current study leverages the enhanced capabilities of both DMRG and advanced FAMPS algorithms, enabling us to accurately simulate systems with sizes up to $L=14$ and perform more reliable analyses of the finite-size scaling behavior of the level-crossing points. Fig.~\ref{crossing} presents the excited-level crossing points $J_2^{c_1}$ and $J_2^{c_2}$ as a function of $1/L$. The excited-level crossing points for $L\leq 10$ are consistent between the previous work \cite{PhysRevLett.121.107202} and our results. 
Remarkably, it is clear from the data that the crossing points scale as $1/L$ instead of the previously assumed $1/L^2$ \cite{10.1063/1.3518900,PhysRevB.94.144416}. {In Fig.~\ref{crossing} (b), we also show a plot of crossing points versus $1/L^2$ for the same data, which clearly deviates from a straight line.}

Extrapolating the level-crossing points through linear fits yields $J_2^{c_1}=0.537(4)$ and $J_2^{c_2}=0.533(1)$, suggesting a direct transition between the N\'eel AFM and VBS phases, ruling out the presence of an intermediate QSL phase. We also perform the same linear fit of the level-crossing points for systems on torus accurately calculated by Y. Nomura, et al. in \cite{PhysRevX.11.031034}, which gives $J_2^{c_1} =0.5344(7)$ and $J_2^{c_2} = 0.538(2)$, consistent with our results.

{Furthermore, we also show the difference of the two crossing point, $\Delta_{J_c}=J_2^{c_2}-J_2^{c_1}$, as a function of $1/L$ in Fig.~\ref{crossing} (c) which goes to zero linearly with $1/L$, indicating the absence of an intermediate spin liquid phase.}

In Fig.~\ref{demo_crossing}, we show how to determine the crossing points of excitation levels by taking the $10 \times 20$ cylinder system as an example. Results for other sizes can be found in the supplementary materials. In Fig.~\ref{demo_crossing} (a), we plot the energies for the singlet and quintuplet excitations for $J_2 = 0.42$ and $0.43$. With the extrapolations of truncation errors, we obtain the accurate energies of the singlet and quintuplet excitations and the difference between them as shown in Fig.~\ref{demo_crossing} (c). Then a linear interpolation gives the crossing point (i.e., the point where $E_2-E_0 = 0$). Using a similar procedure, we determine the crossing point for singlet and triplet excitations, as depicted in panels (b) and (d) of Fig.~\ref{demo_crossing}.  

In the supplementary materials, we calculate the staggered magnetization of the N\'eel AFM order at $J_2 = 0.5$, which is clearly nonzero (about $0.05$). This result show $J_2 = 0.5$ is in the N\'eel AFM phase. The boundary of the N\'eel AFM phase obtained from the finite size scaling of the crossing point of correlation length is $J_2 = 0.530$ (details can be found in the supplementary materials), consistent with the results from level-crossing analysis.  

\begin{figure}[t]
	\includegraphics[width=85mm]{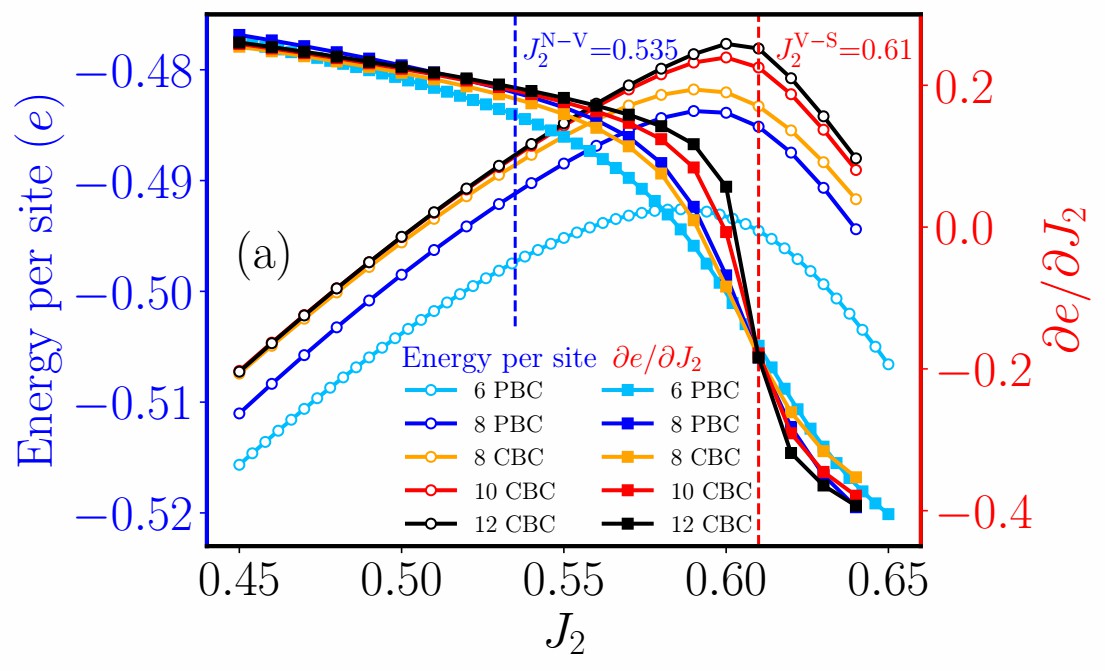}
   \hbox{\hspace{-7mm} \includegraphics[width=71mm]{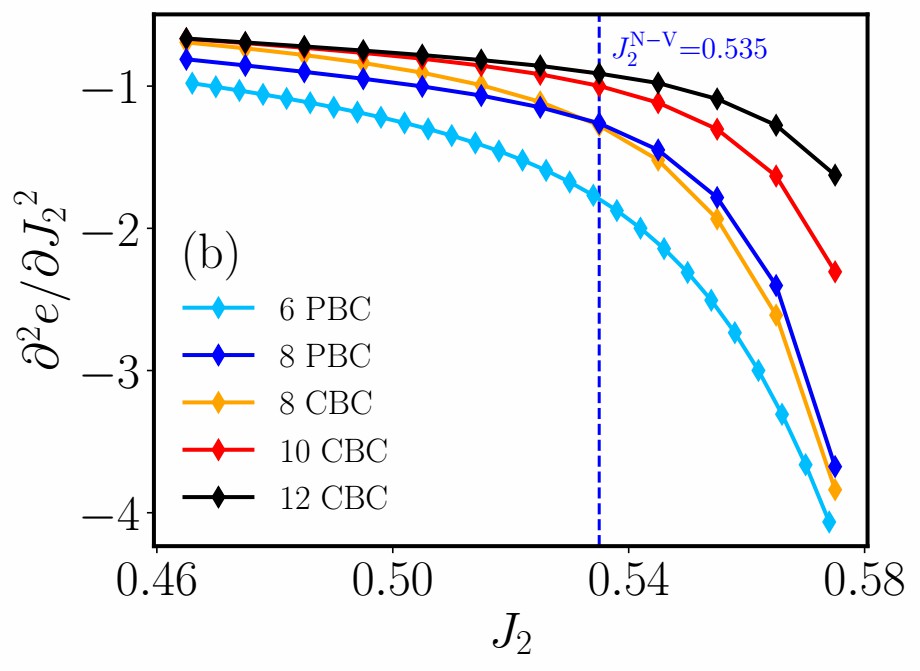}}
   \caption{{ (a) The ground state energy per site $e$ and its first derivative $\partial e/\partial J_2$ as a function of $J_2$ for $L\times L$ systems with periodic boundary conditions (PBC) and $L\times 2L$ systems with cylinder boundary conditions (CBC). Here, we find the phase transition between N\'eel AFM and VBS phases at $J_2^{\text{N-V}}=0.535$ is continuous. The crossing point of $\partial e/\partial J_2$ gives a precise location of the first order transition between VBS and stripe AFM phases at $J_2^{\text{V-S}}=0.610$ (see the text for more discussions). (b) The second derivative of the ground state energy as a function of $J_2$ for different systems. The calculations are performed using DMRG, with maximum of equivalent $60000$ U(1) states retained, except for $6\times6$ PBC systems which are calculated by exact diagonalization \cite{10.21468/SciPostPhys.7.2.020}.}
   }
   \label{energy}
\end{figure}

{\em \Y{Energitics --} } We also study the behavior of ground state energy versus $J_2$ to detect the quantum phase transitions. Fig.~\ref{energy} (a) shows the ground state energy density $e$ and its first derivative $\partial e/\partial J_2$ as a function of $J_2$. We calculate $\partial e/\partial J_2$ from the expectation value of the $J_2$ term in the ground state according to the Feynman-Hellman theorem.  We show results for both torus systems ($6 \times 6$ and $8 \times 8$) and cylinder systems ($8 \times 16$, $10 \times 20$, and $12 \times 24$). Here we only show the results with the largest bond dimensions reached (the convergence of these results with bond dimensions can be found in the supplementary materials). 

At the aforementioned critical point, $J_2^{\text{N-V}}=0.535$, it is intriguing to note that $\partial e/\partial J_2$ is continuous. This behavior strongly suggests the transition between the N\'eel AFM and the VBS phases is continuous. This observation aligns with the concept of DQCP \cite{doi:10.1126/science.1091806,senthil2023deconfined}, which is proposed to elucidate the continuous phase transition between the N\'eel AFM state and the VBS state. We notice that the deconfined criticality near $J_2=0.54$ was claimed before \cite{PhysRevX.12.031039,PhysRevX.11.031034,liu2023emergent,PhysRevB.97.174408,PhysRevLett.113.027201,PhysRevB.94.075143}. We also calculate the second derivative of the ground state energy with respect to $J_2$, i.e., $\partial^2 e/ {\partial J_2}^2$, using the finite difference method, as shown in Fig.~\ref{energy} (b).
Interestingly, there is no tendency of singularity at the critical point between the N\'eel AFM and the VBS phases near $J_2^{\text{N-V}}=0.535$ in $\partial^2 e/ {\partial J_2}^2$, suggesting the phase transition is a high-order one. Further characterization of this phase transition requires additional investigations. 

Finally, We shift our focus to the easier segment of the phase diagram. It was established that there exists a first order phase transition between the VBS and stripe AFM phases at $J_2 \approx 0.6$ \cite{PhysRevLett.63.2148,QMC_Imada}. In Fig.~\ref{energy} (a), we can clearly find a discontinuity in the $\partial e/\partial J_2$ plot near $J_2 = 0.6$. As elucidated in Ref.~\cite{1990JSP6179B}, for a first-order transition, finite size scaling analysis reveals a distinct point denoted as $h(L)$. At this point $h(L)$, the quantity $M_{\text{per}}(h(L),L)$ remains constant with the varying of system size $L$. Furthermore, the difference between $h(L)$ and $h_c$ is bounded by $O(e^{-L})$. Here, $h$ represents the parameter driving the phase transition, $h_c$ is the critical point at the thermodynamic limit, while $M_{\text{per}}$ corresponds to either the order parameter or the first derivative of the free energy density. The exponentially small difference between $h(L)$ and $h_c$ suggests the existence of a fixed point for the first derivative of the free energy density and allows us to accurately determine the location of the first order transition. 
In Fig.~\ref{energy} (a), we can clearly find a fixed point in $\partial e/\partial J_2$ at $J_2 = 0.610$. We thus conclude that the first order transition between the VBS and stripe AFM phases occurs at $J_2=0.610(5)$.

{\em \Y{Conclusions --}}
With accurate DMRG and FAMPS results and careful finite size scaling of the excited-level crossing points, we demonstrate a direct phase transition between the N\'eel AFM and the VBS phases at $J_2^{\text{N-V}}=0.535(3)$ for the $J_1-J_2$ Heisenberg model on the square lattice, indicating the absence of the previously claimed intermediate quantum spin liquid phases \cite{PhysRevB.88.060402,PhysRevB.86.024424,PhysRevLett.121.107202,PhysRevX.11.031034,LIU20221034}. Moreover, the results from energy show the phase transition is continuous, suggesting a deconfined quantum critical point at $J_2^{\text{N-V}}=0.535(3)$ which deserves further explorations. We also determine the precise location of the first order phase transition between the VBS and stripe phases at $J_2^{\text{V-S}}=0.610(5)$ from the crossing of the first derivative of energy for different system sizes.




Looking ahead, the absence of a spin liquid phase in the $J_1-J_2$ Heisenberg model on the square lattice prompts further inquiries into the roles of additional interactions and lattice geometries in shaping the behavior of quantum materials \cite{PhysRevX.12.031039,liu2023emergent,doi:10.1126/science.1201080,PhysRevLett.109.067201,PhysRevLett.118.137202}. 

\begin{acknowledgments}
The calculation in this work is carried out with TensorKit \cite{foot2}. {The computation in this paper were run on the Siyuan-1 cluster supported by the Center for High Performance Computing at Shanghai Jiao Tong University.} MQ acknowledges the
support from the National Key Research and Development Program of MOST of China (2022YFA1405400), the Innovation Program for Quantum Science and Technology (2021ZD0301902),
the National Natural Science Foundation of China (Grant
No. 12274290) and the sponsorship from Yangyang Development Fund.
\end{acknowledgments}

\bibliography{Heisenberg}

\begin{thebibliography}{83}
\bibitem[{foo({\natexlab{a}})}]{foot1}
\bibinfo{note}{For systems with a width of $L \leq 12$, we leverage DMRG
  method, preserving up to 60000 U(1) states to ensure accurate results. In the
  case of larger system of $L=14$, we use FAMPS to perform the calculation,
  where we keep a maximum of 22000 U(1) states.}
\bibitem[{foo({\natexlab{b}})}]{foot3}
\bibinfo{note}{Disentanglers are unitary matrices which transform the physical
  degrees of freedom of two sites. They are common building blocks in Tensor
  Networks which can reduce the local entanglement in the studied system and
  helps to improve the accuracy of the simulation
  \cite{PhysRevLett.101.110501}}.
\end{thebibliography}

\end{document}


\title{Supplementary Materials for ``Absence of Spin Liquid Phase in the $J_1-J_2$ Heisenberg model on the Square Lattice''}

\author{Xiangjian Qian}
\affiliation{Key Laboratory of Artificial Structures and Quantum Control (Ministry of Education),  School of Physics and Astronomy, Shanghai Jiao Tong University, Shanghai 200240, China}

\author{Mingpu Qin} \thanks{qinmingpu@sjtu.edu.cn}
\affiliation{Key Laboratory of Artificial Structures and Quantum Control (Ministry of Education),  School of Physics and Astronomy, Shanghai Jiao Tong University, Shanghai 200240, China}

\affiliation{Hefei National Laboratory, Hefei 230088, China}

\maketitle

\begin{figure*}[t]
	\includegraphics[width=85mm]{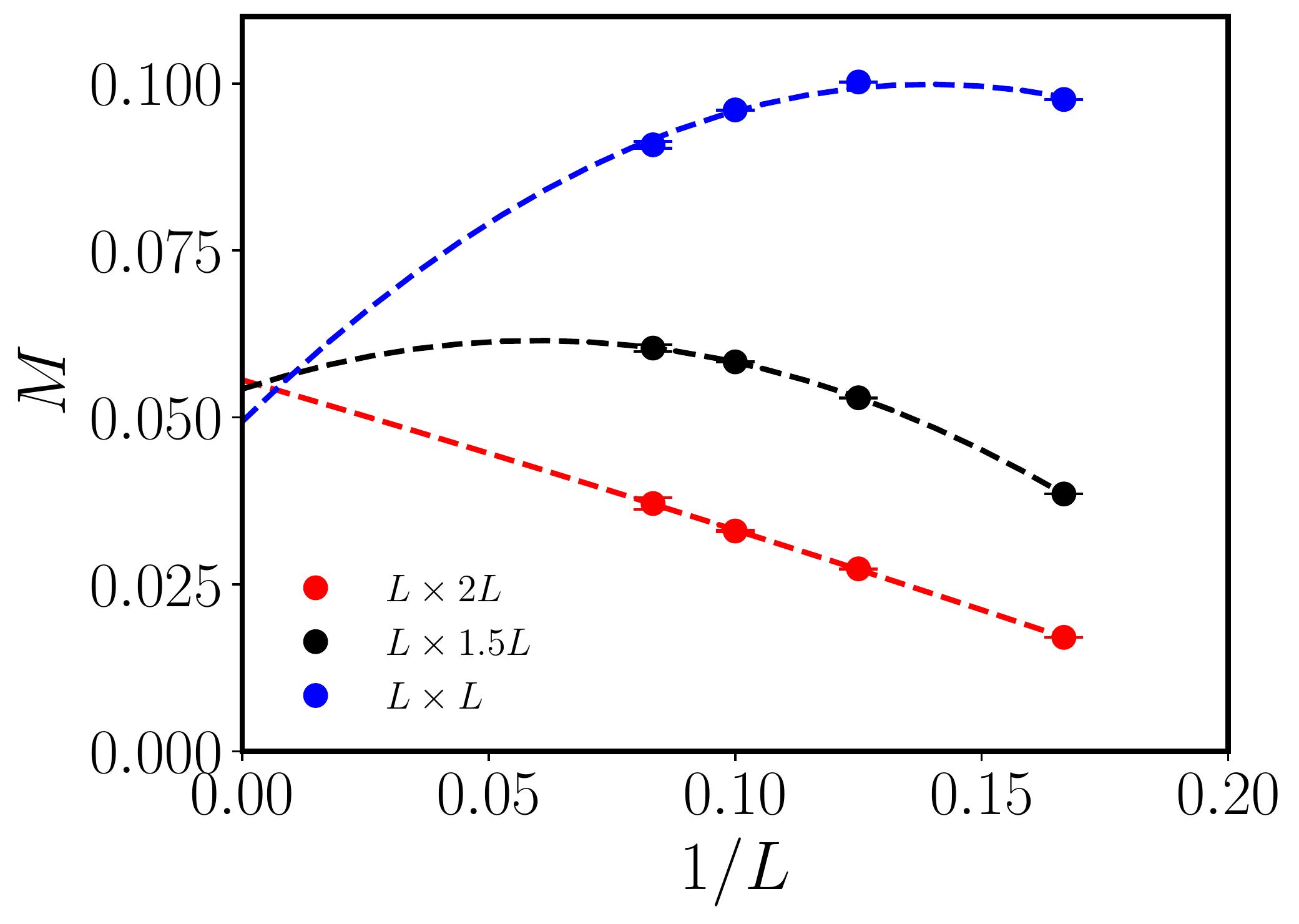}
	\caption{{$M$ vs $1/L$ results for the $J_1-J_2$ Heisenberg model on the square lattice at $J_2=0.5$. The simulations are performed on $L\times rL$ (with aspect ratio r) cylinder systems with pinning fields on the open edges as shown in \cite{PhysRevLett.99.127004}. An estimation of N\'eel AFM order is $M\approx 0.05$.}}
	\label{AFM_phase}
\end{figure*}

\begin{figure*}[t]
	\includegraphics[width=85mm]{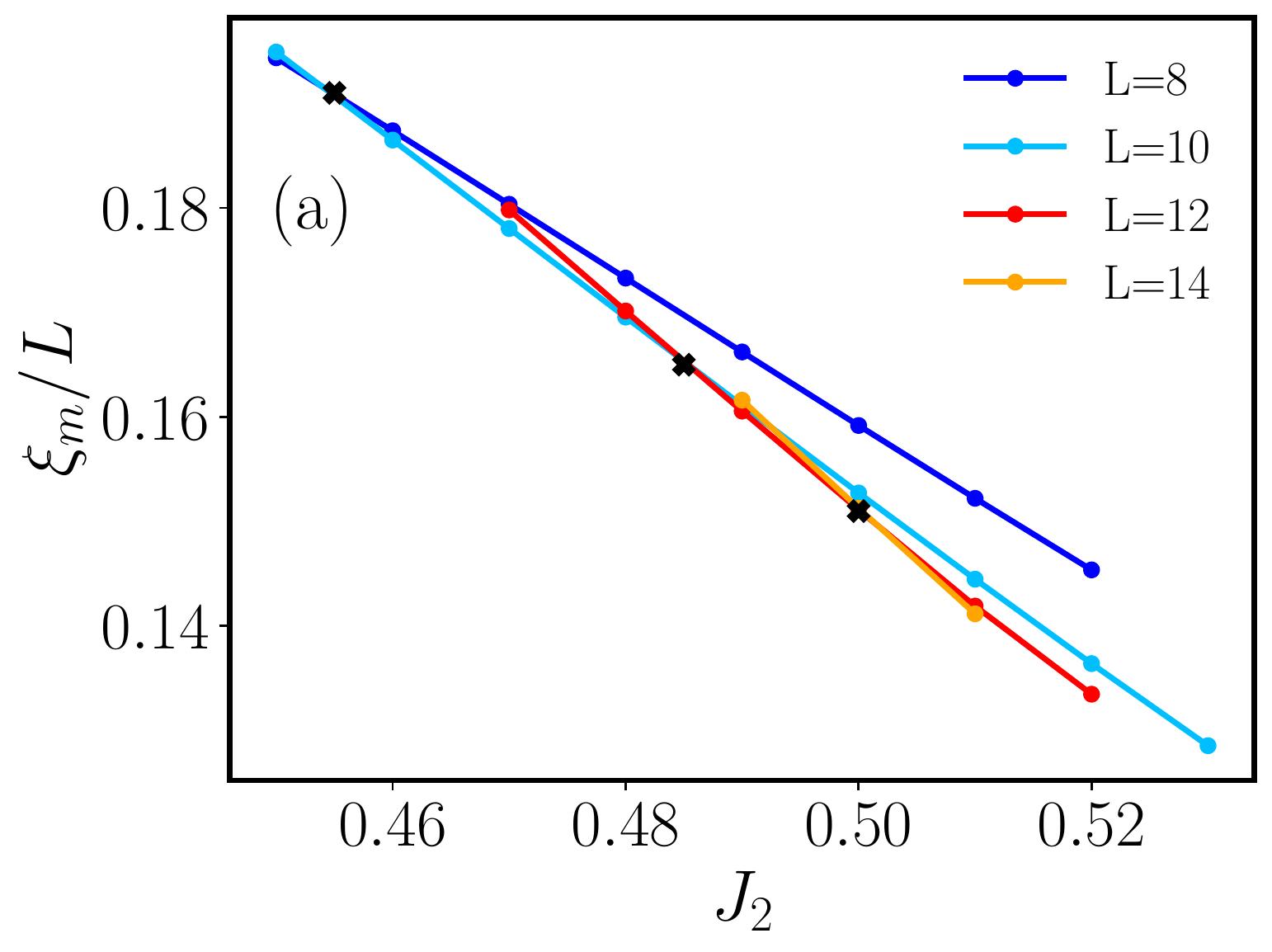}
	\includegraphics[width=85mm]{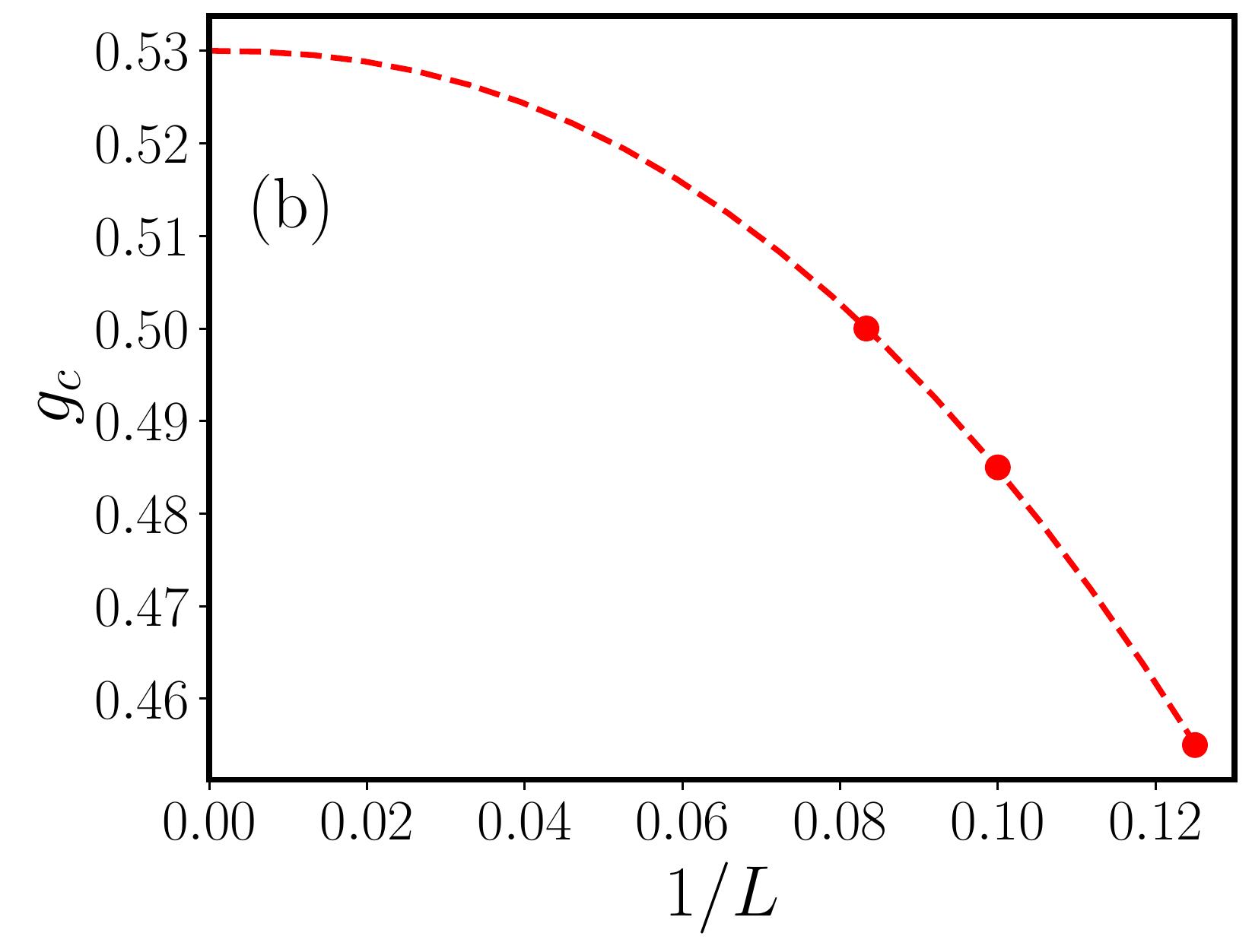}
	\caption{{(a) $\xi_m/L$ as a function of $J_2$ for different system sizes.  (b) Crossing points $g_c$ of $\xi_m/L$ as a function of $1/L$. A power law fit $(a+bL^{-w})$ yields $J_2^{c_1}=0.530$, consistent with the phase boundary of AFM phase we obtain from the crossing of excitation gaps.} }
	\label{CORR}
\end{figure*}

\subsection{N\'eel AFM order}
{Since there is a consensus on the onset of the VBS phase at $J_2\approx 0.54$ \cite{PhysRevX.11.031034, PhysRevB.97.174408, PhysRevLett.113.027201, PhysRevLett.121.107202}, we focus on the determination of the boundary of the N\'eel phase. }

{Firstly, we consider the N\'eel AFM order at $J_2=0.5$ which is deep inside the N\'eel AFM phase in our phase diagram and is considered to be the most frustrated point in the $J_1-J_2$ Heisenberg model. Precious investigations proposed the existence of a spin liquid state at $J_2=0.5$ \cite{PhysRevLett.121.107202,PhysRevLett.113.027201,LIU20221034,PhysRevX.11.031034}, suggesting the absence of N\'eel AFM order at $J_2=0.5$. However, due to the limitation on the system sizes we can accurately simulate with DMRG, the task of precisely identifying the magnetization in the thermodynamic limit via finite scaling analysis is challenging \cite{LIU20221034,PhysRevX.11.031034}. In contrast, we adopt a more accurate strategy employed in Ref \cite{PhysRevLett.99.127004} to detect the N\'eel AFM order. We add AF pinning fields in the open edges of the system to detect the AFM order parameter in the thermodynamic limit,
from the finite size scaling of the bulk order parameter for systems with different aspect ratios. Our results are summarized in Fig.~\ref{AFM_phase}. We can find that the staggered magnetization is converged in the thermodynamic limit for systems with aspect ratios $1,1.5$, and $2$. The AFM order parameter $M$ is estimated to be around around $0.05$ at $J_2=0.5$, which means $J_2=0.5$ is in the AFM phase. This result provides another strong evidence to support the phase diagram we obtain.}

{Secondly, to determine the boundary of the N\'eel AFM phase, we calculate a dimensionless quantity $\xi_m/L$ ($\xi_m$ is the correlation length) which is defined as:
\begin{equation}
	\xi_m/L=\frac{1}{2\pi}\sqrt{\frac{m_s^2(\pi,\pi)}{m_s^2(\pi+2\pi/L,\pi)}-1}
\end{equation}
where $m_s^2(\bm{k})=\frac{1}{L^4} \sum_{\bm{ij}}\langle S_{\bm{i}}\cdot S_{\bm{j}} \rangle e^{i\bm{k}(\bm{i}-\bm{j})}$, $\bm{i}=(i_x,i_y)$ is the site position and the summation runs over the center $L\times L$ part of the $L\times 2L$ system. $\xi_m/L$ diverges in the N\'eel AFM phase and is $0$ in a disordered phase. On finite systems, this quantity shows a crossing point $g_c$ near the critical point making it a valuable tool for detecting phase transition point \cite{10.1063/1.3518900,PhysRevLett.115.157202}. Our results are shown in Fig.~\ref{CORR}. We find the crossing point $g_c$ shift a lot with the system size. The crossing points between $L=12$ and $L=14$ is already larger than $J_2=0.5$, consistent with our finding of nonzero N\'eel AFM order at $J_2=0.5$. A simple power law fit ($a+bL^{-w}$) for $g_c$ yields $J_2^{c_1}=0.530$ \cite{doi:10.1126/science.aad5007}, consistent with the phase boundary of AFM phase we obtain from the crossing of excitation gaps.}

\begin{figure*}[t]
	\includegraphics[width=85mm]{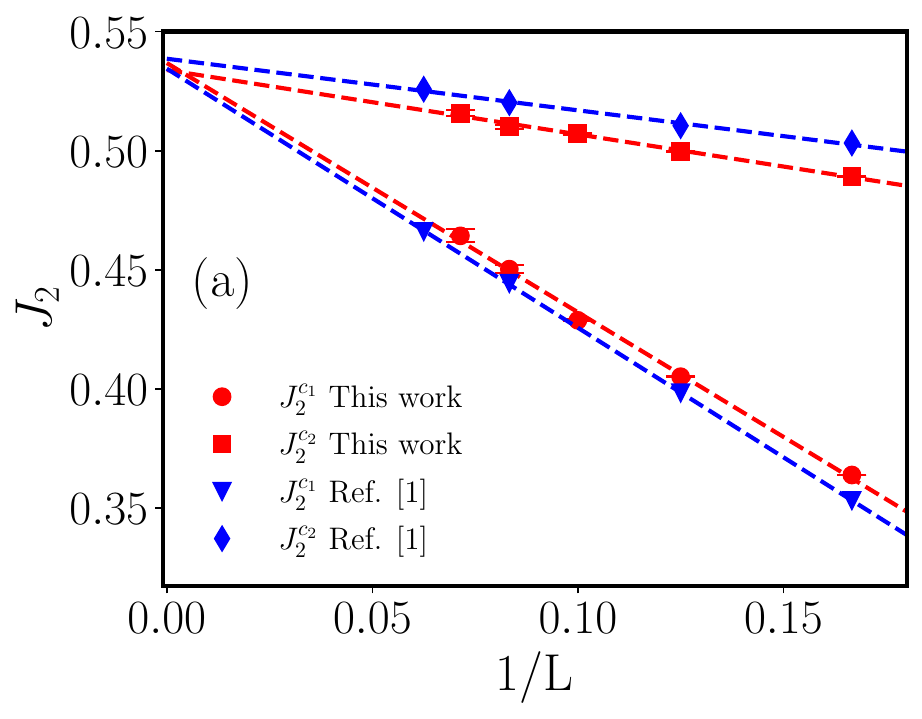}
	\includegraphics[width=85mm]{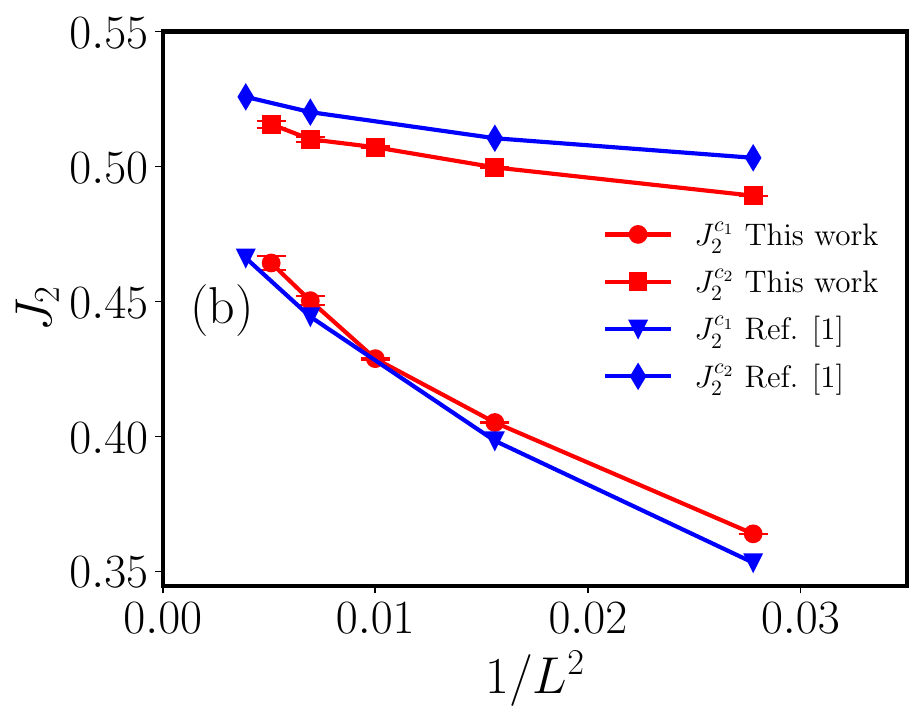}
	\caption{(a) The linear fits with respect to $1/L$ of the excited level crossing points. Both of the data from the torus \cite{PhysRevX.11.031034} and cylinder systems converge to the same points $J_2\approx 0.535$. Specifically, our extrapolated values are $J_2^{c_1}=0.537(4)$ and $J_2^{c_2}=0.533(1)$. The extrapolated results from the data in \cite{PhysRevX.11.031034} are $J_2^{c_1} =0.5344(7)$ and $J_2^{c_2} = 0.538(2)$. (b) We plot the same data in (a) by changing x axis to $1/L^2$. We can clearly see the deviation from a straight line both in torus systems and cylinder systems.}
	\label{crossing_points}
\end{figure*}

\subsection{More discussions about the scaling behavior of the excited energy crossing points}

Firstly, we present a comparison between our cylinder results for the excited crossing points and the accurate torus results from a previous study \cite{PhysRevX.11.031034}, as depicted in Fig.~\ref{crossing_points}. Both sets of results exhibit a linear dependence on $1/L$ for the excited crossing points. Specifically, our linearly extrapolated values are $J_2^{c_1}=0.537(4)$ and $J_2^{c_2}=0.533(1)$. The extrapolated results from the data in Ref. \cite{PhysRevX.11.031034} are $J_2^{c_1} =0.5344(7)$ and $J_2^{c_2} = 0.538(2)$, consistent with our results. We also show the crossing points as a function of $1/L^2$ in Fig.~\ref{crossing_points} (b), in which we can find an obvious derivation from a straight line both in torus and cylinder systems.


\subsection{Convergence of the energies}
\begin{figure*}[t]
	\includegraphics[width=85mm]{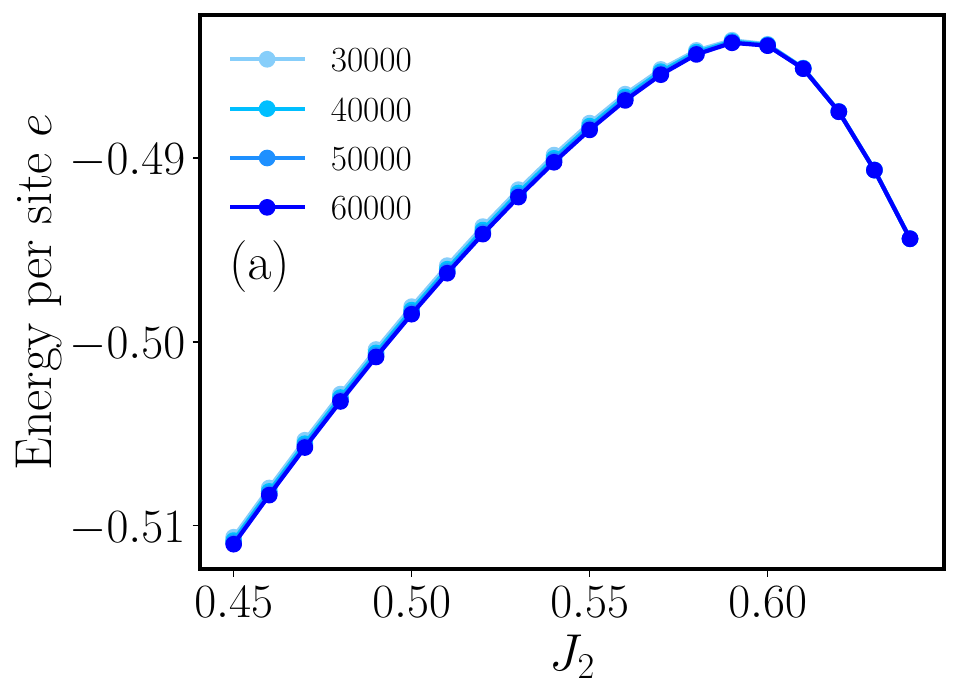}
	\includegraphics[width=85mm]{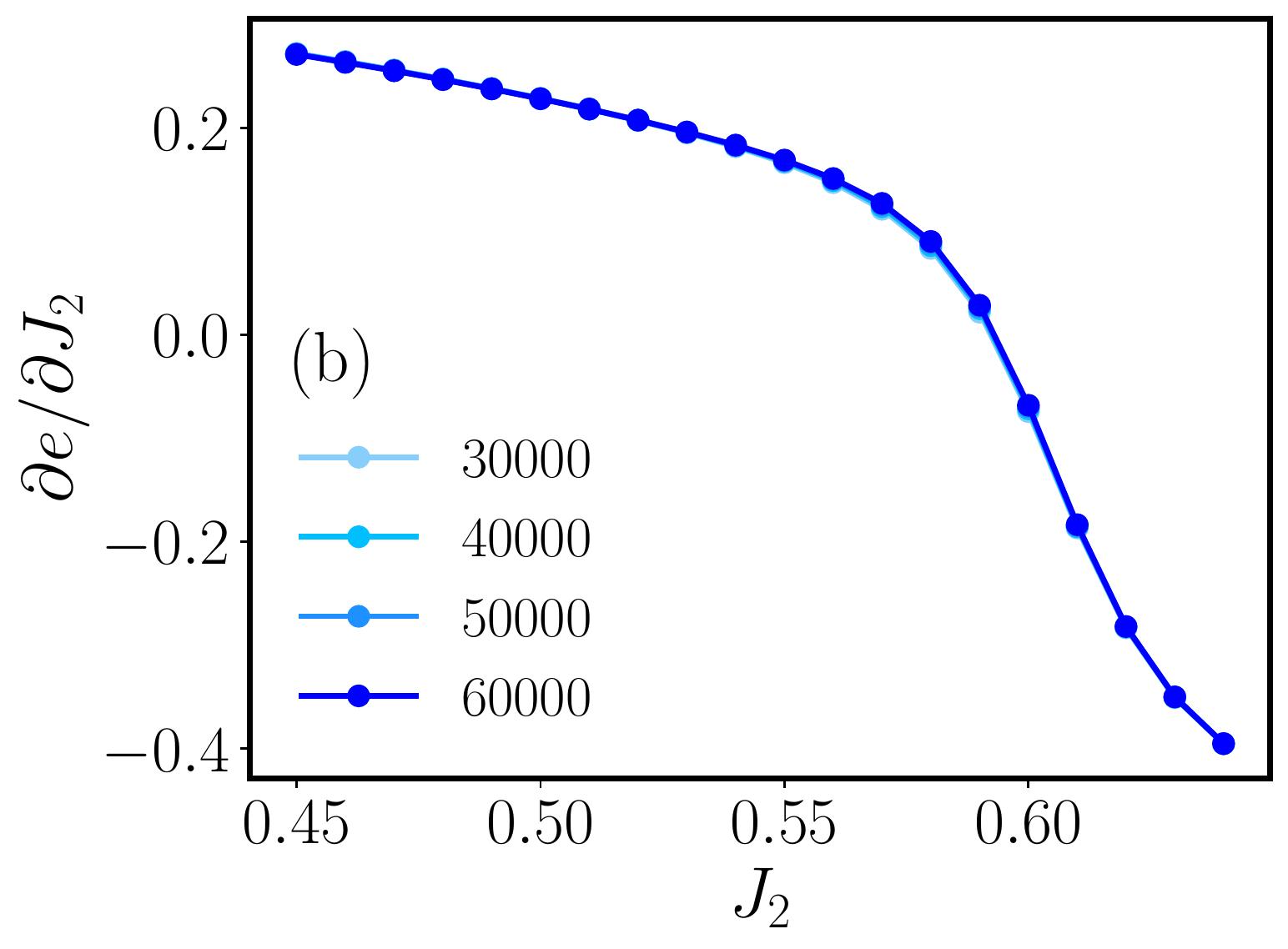}
	\includegraphics[width=85mm]{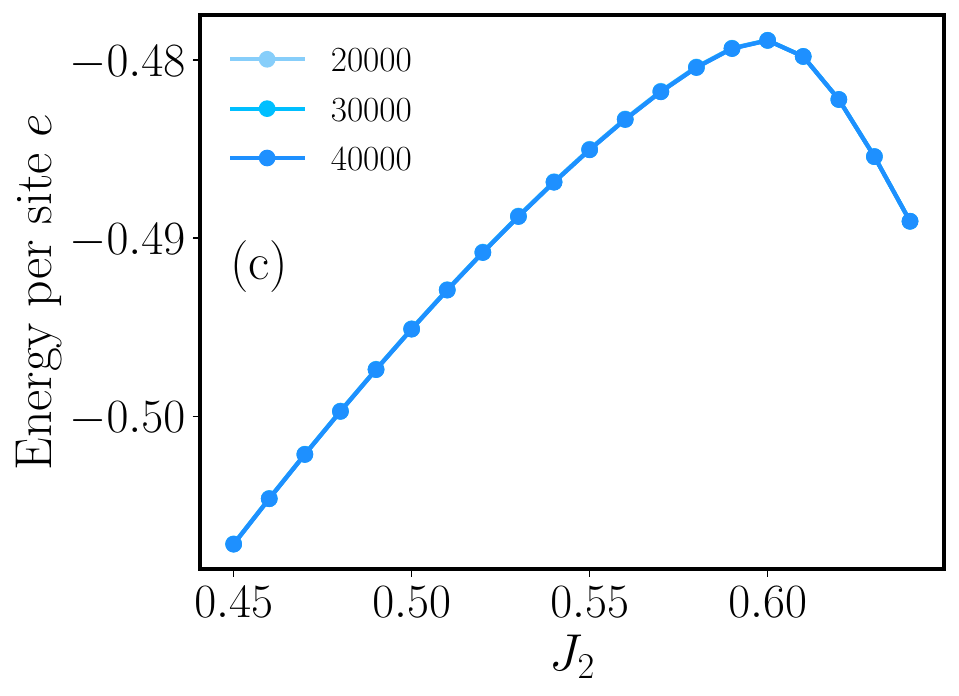}
	\includegraphics[width=85mm]{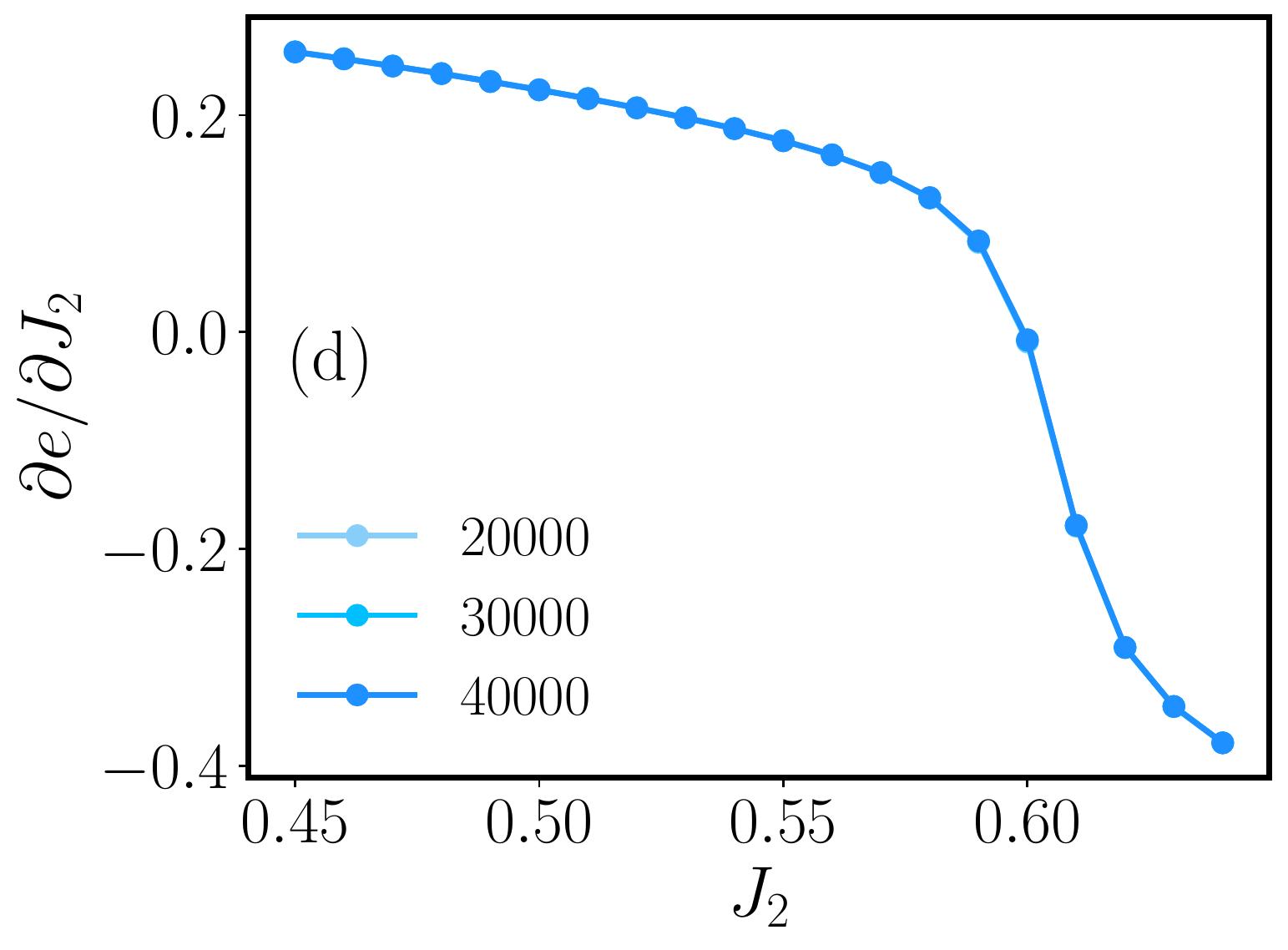}
	\includegraphics[width=85mm]{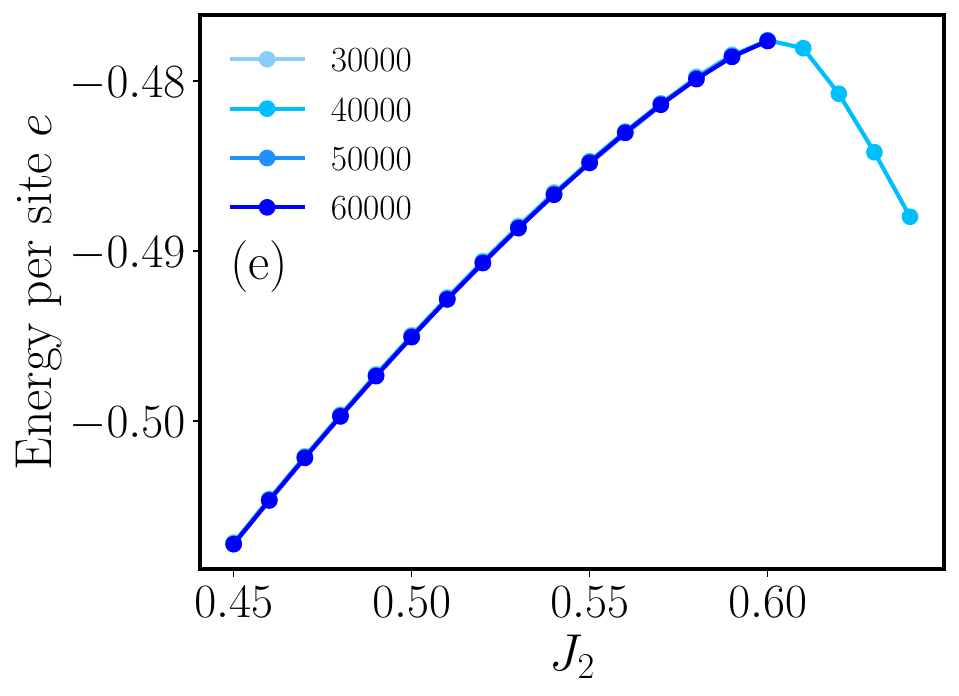}
	\includegraphics[width=85mm]{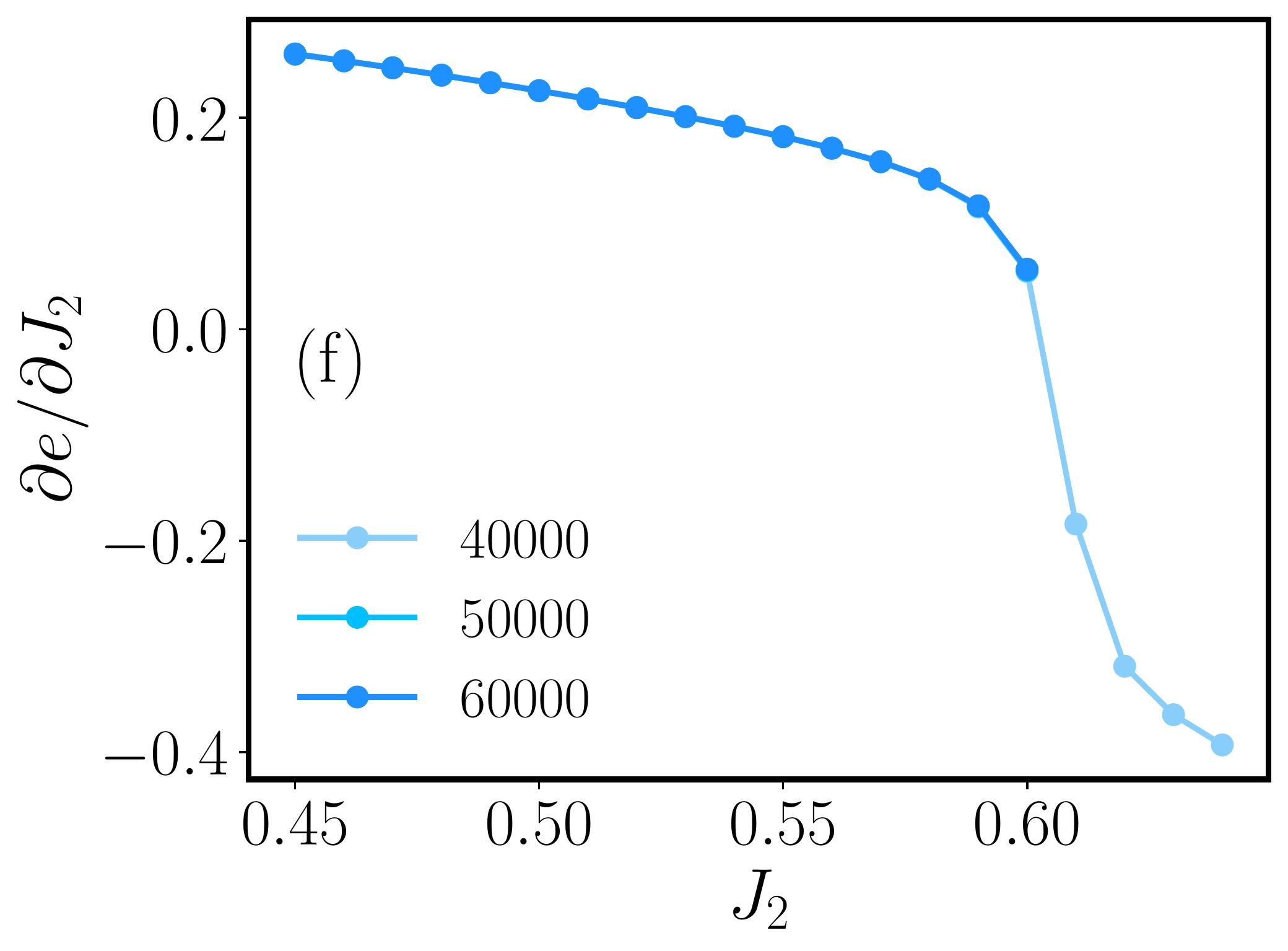}
	\caption{The ground state energy per site and its first derivative as a function of $J_2$ for different bond dimensions. (a), (b) Results for a $8\times 8$ torus system. (c), (d) Results for a $10\times 20$ cylinder system (e), (f) results for a $12\times 24$ cylinder system.}
	\label{ground_state}
\end{figure*}

In Fig.~\ref{ground_state}, we show the convergence of the ground state energy and its derivative with respect to $J_2$ for different system sizes.

\subsection{Extrapolation of the excitation levels}
\begin{figure*}[t]
	\includegraphics[width=85mm]{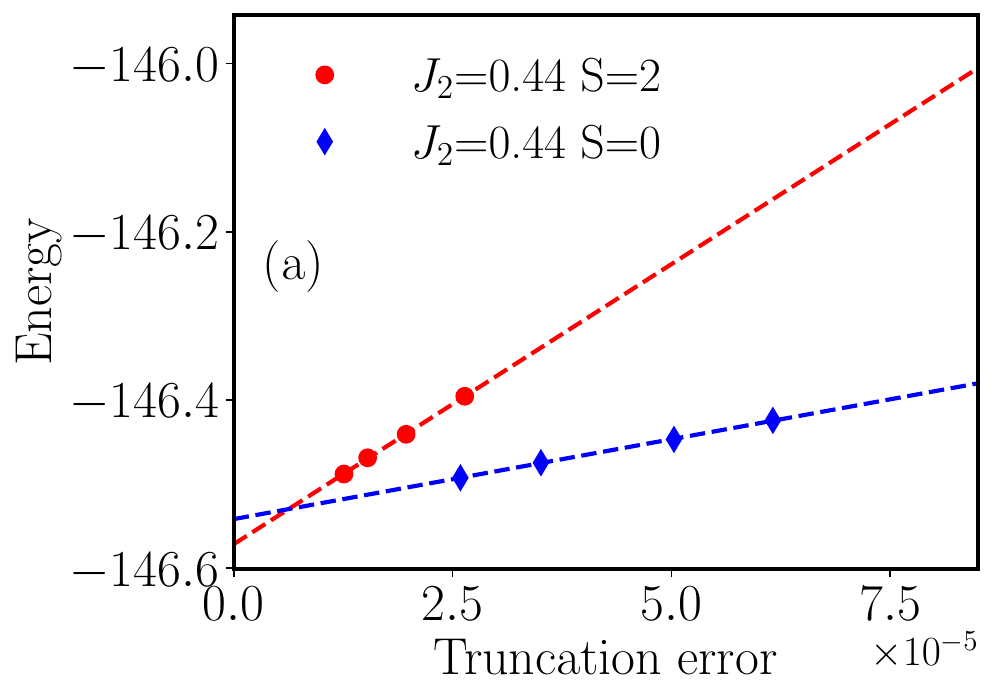}
	\includegraphics[width=85mm]{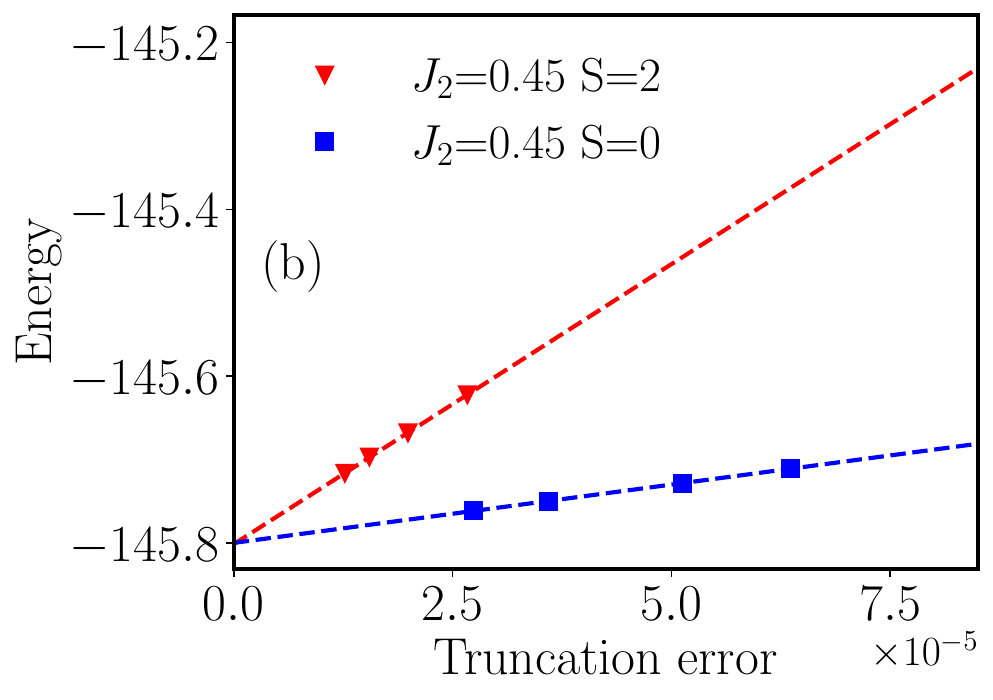}
	\includegraphics[width=85mm]{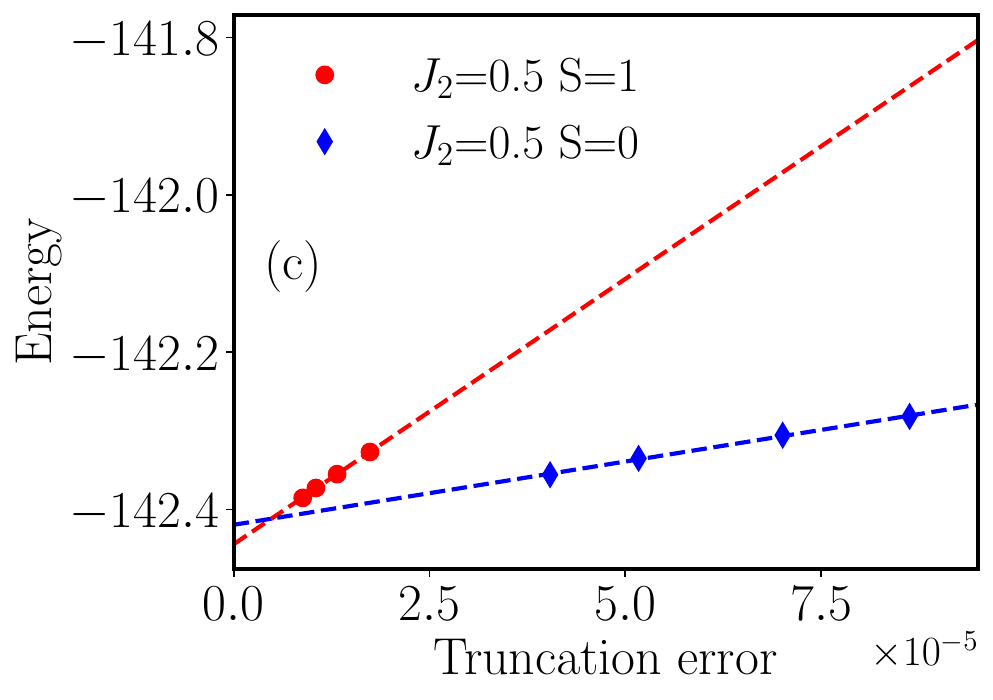}
	\includegraphics[width=85mm]{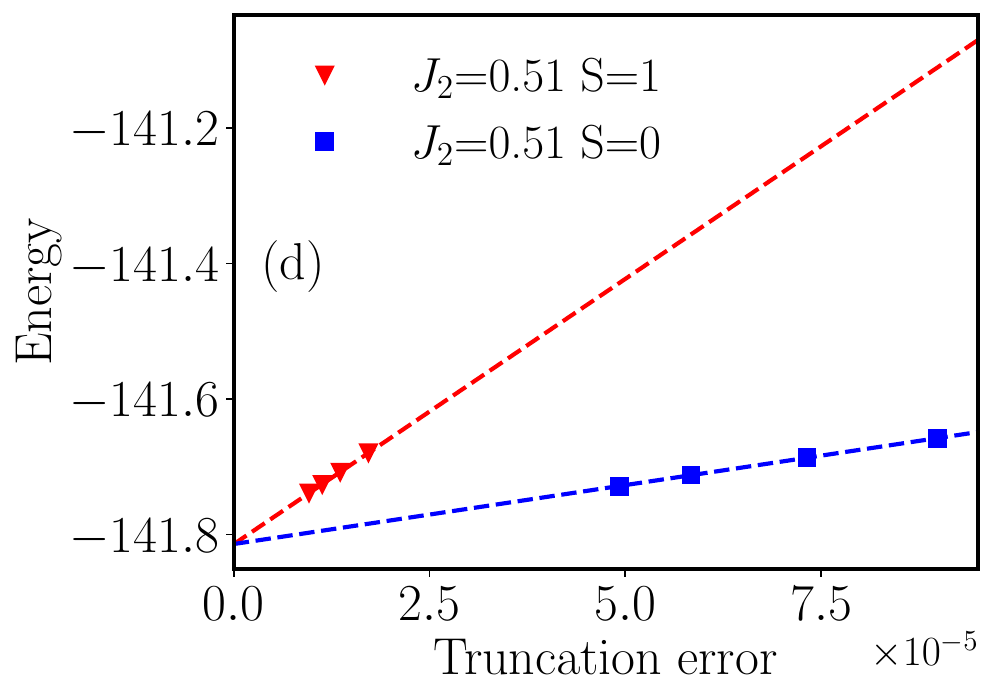}
	\includegraphics[width=85mm]{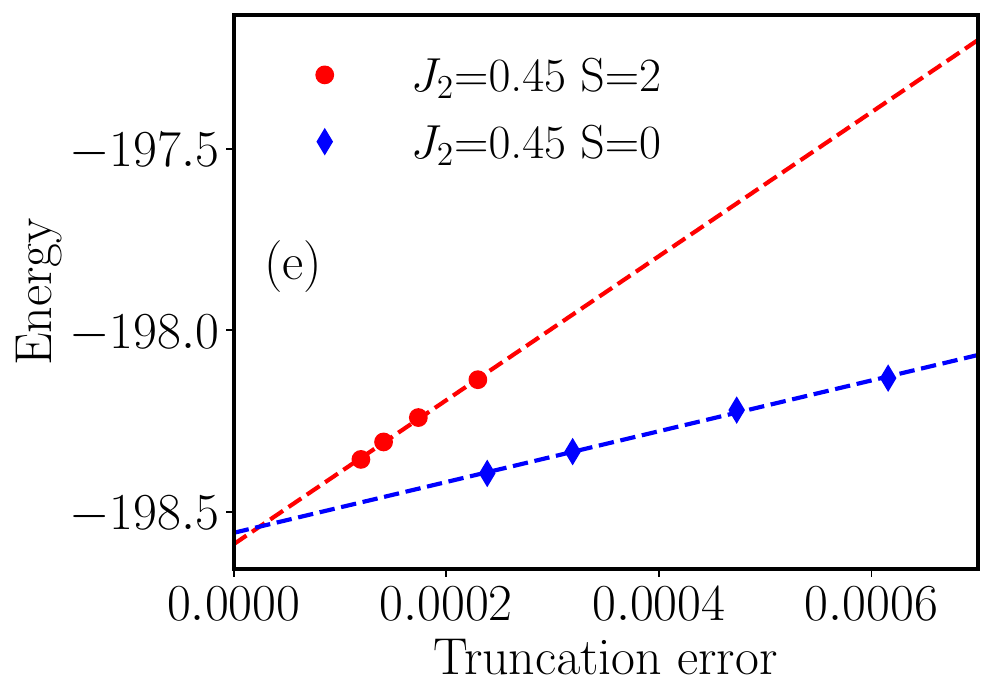}
	\includegraphics[width=85mm]{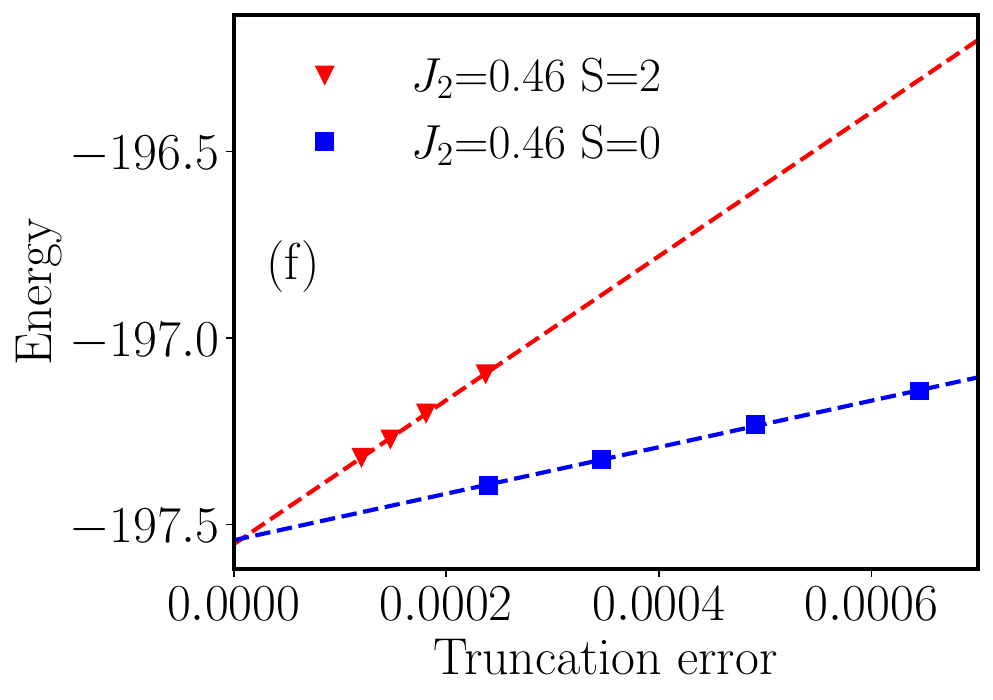}
 	\includegraphics[width=85mm]{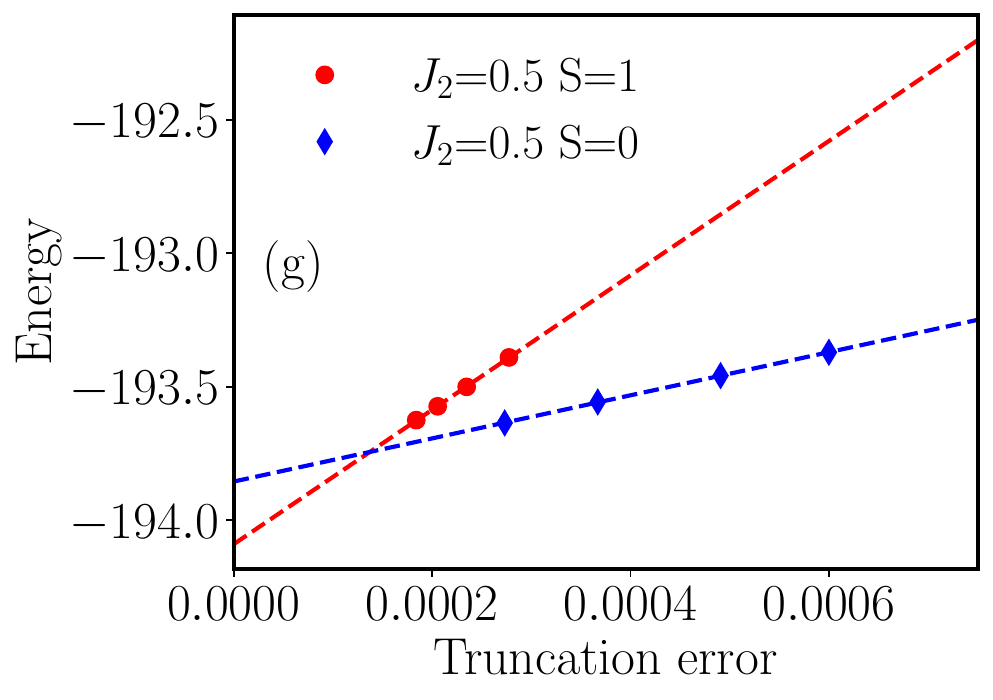}
	\includegraphics[width=85mm]{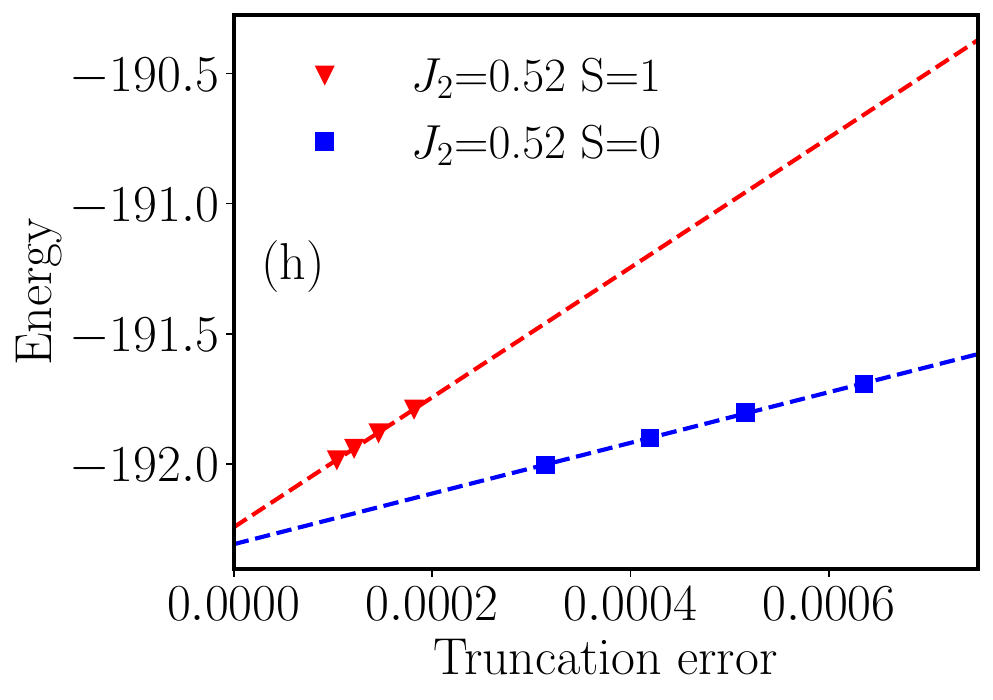}
	\caption{(a)-(d) Energies of the relevant $S=0, S=1$ and $S=2$ excitations for the $12\times 24$ cylinder system. (e)-(h) Energies of the relevant $S=0, S=1$ and $S=2$ excitations for the $14\times 28$ cylinder system. Extrapolations with truncation errors are shown for all energies.}
	\label{extra_Hei}
\end{figure*}

In Fig.~\ref{extra_Hei}, we show the extrapolation of the energies of the excited state with truncation errors for different systems. For the $12\times 24$ cylinder system, we perform the calculation using DMRG, with a maximal 60000 U(1) equivalent kept states. For $14\times 28$ systems, we employ FAMPS by keeping maximal 22000 U(1) equivalent states. 
In FAMPS calculations we block two sites into one \cite{Qian_2023}, which results in larger truncation errors than in conventional DMRG calculation. 

\bibliography{Heisenberg_SM}